%% file: main.tex
\input{preambule.tex}

\begin{document}
\input{Methode}
\addcontentsline{toc}{chapter}{References}
\nocite{*}
\renewcommand{\refname}{References}
\bibliographystyle{apalike}
\bibliography{References.bib}

\end{document}

%% file: preambule.tex
\documentclass[12pt]{article}

\usepackage{tikz}
\usetikzlibrary{positioning}
\usetikzlibrary{calc}
\usepackage{subcaption}

\usepackage{pstricks}
\usepackage{pst-tree}
\usepackage[]{graphicx}
\usepackage[]{color}
\usepackage[T1]{fontenc}
\usepackage[utf8]{inputenc}
\usepackage{geometry}
\usepackage[numbers]{natbib}
\usepackage[nottoc]{tocbibind}
\usepackage{float}
\usepackage{soul}
\usepackage{amsmath}
\usepackage{amssymb}
\usepackage{setspace}
\usepackage{amsfonts}
\usepackage{dsfont}
\usepackage[english]{babel}
\usepackage[labelfont = {bf}, 
  textfont ={it}, 
  margin =1cm,
  font=small, 
  tablename = Table ]{caption}
  
\usepackage[labelfont = {bf}, 
  textfont ={it}, 
  margin =1cm,
  font=small, 
  figurename = Figure ]{caption}
  
\addto\captionsfrench{
  
}

\usepackage{amsthm} 
\usepackage{bbm}
\usepackage{booktabs}
\usepackage{float}
\usepackage{colortbl}
\usepackage{dcolumn}
\usepackage{icomma}
\usepackage{fancyhdr}
\usepackage{titlesec}
\titleformat{\section}
{\Huge\bfseries}{\thesection}{12pt}{}
\titleformat{\subsection}
{\large\bfseries}{\thesubsection}{12pt}{}[]
\titleformat{\subsubsection}
{\bfseries}{\thesubsubsection}{12pt}{}[]
\usepackage{listings}
\usepackage[flushleft]{threeparttable}
\usepackage{caption}
\usepackage[tikz]{bclogo}
\usepackage{pgfornament}
\usepackage{url} 
\usepackage[hidelinks]{hyperref}
\usepackage{orcidlink}
\usepackage{hyperref}
\usepackage{fancyhdr} 
\usepackage{fancybox}
\usepackage{pifont}
\usepackage{multirow}
\usepackage{booktabs}
\usepackage{makecell}

\renewcommand{\refname}{References}

%% file: Methode.tex
\begin{titlepage}
\newcommand{\HRule}{\rule{\linewidth}{0.5mm}} 
\begin{center}
  
 \textbf{Risk prediction models for breast cancer integrating family history of breast cancer and prophylactic interventions}
\end{center}

\vspace{0.5cm}
\begin{center}
   Siriac Seboka\textsuperscript{1}, Laurent Briollais\textsuperscript{2,3}, Yun-Hee Choi\textsuperscript{4}, on behalf of BCFR,
Lajmi Lakhal-Chaieb \orcidlink{https://orcid.org/0000-0002-8358-4392}\textsuperscript{1,*}\\

 {\small \textsuperscript{1}Département de Mathématiques et de Statistique, Université Laval, Québec, Canada}\\{\small \textsuperscript{2}Lunenfeld-Tanenbaum Research Institute, Mount Sinai Hospital, Toronto, Canada}\\
 {\small \textsuperscript{3}Biostatistics division, Dalla Lana School of Public Health, University of Toronto, Toronto, Canada}\\
 {\small \textsuperscript{4}Department of Epidemiology and Biostatistics, Western University, London, Canada}\\
{\small \textsuperscript{*}Correspondence: Lajmi Lakal-Chaieb (lakhal@mat.ulaval.ca)} 
\end{center}


\noindent \textbf{Abstract}:
\textcolor{black}{Breast cancer risk assessment and prediction models are very important tools for the clinical management of healthy women carrying pathogenic variants in known cancer genes. We propose in this paper a comprehensive prediction model for the personalized clinical management of women with pathogenic variants in {\it BRCA1}. This model is able to estimate the risk of BC accounting for the full history of this cancer within the family, the personal and family history of risk-reducing salpingo-oophorectomy as well as the exact age of this intervention among family members. The effect of oophorectomy on the risk of developing breast cancer is evaluated by estimating and conducting inference about the regression coefficients in a Cox model with time-varying covariates. We model the within-family dependence in the ages at onset of cancer using a Gaussian copula whose correlation matrix accommodates the different pairwise family relationships. We develop an iterative algorithm to estimate the parameters of the considered model in the presence of a selection bias. We evaluated the performance of the proposed cancer risk estimation method from family data by simulations and illustrated its use through an application to the breast cancer family registry.}


\end{titlepage}       
\section{Introduction}

\noindent Women carrying pathogenic or likely pathogenic variants (P/LPVs) in {\it BRCA1} and/or {\it BRCA2} ({\it BRCA1/2}) genes are at high risk of developing breast cancer (BC) and ovarian cancer (OC). Large meta-analyses reported cumulative risks of BC by age 70 years reaching 57\% for {\it BRCA1} P/LPV carriers and 49\% for {\it BRCA2} carriers \citep{chen2007meta},\citep{kuchenbaecker2017risks}. The equivalent OC risks are estimated at 40\% for {\it BRCA1} and 18\% for {\it BRCA2} P/LPV carriers. Women with {\it BRCA1} or {\it BRCA2} pathogenic variants undergo intensive cancer surveillance and may also consider surgical interventions such as risk-reducing mastectomy (RRM) or risk-reducing salpingo oophorectomy (RRSO). \textcolor{black}{Cancer risk assessment and prediction models are very important tools for the clinical management of healthy women carrying a {\it BRCA1} or {\it BRCA2} P/LPV, which may involve screening planning, risk-reducing surgeries and chemo-prevention. Risk prediction models for breast and ovarian cancers such as BCRAT \citep{gail1989} can integrate a woman's own personal and family history to calculate the risk of cancer over specific periods of time. Some prediction models are specifically designed to estimate risks for {\it BRCA1/2} PV carriers. For instance, BOADICEA \citep{lee2019} calculates breast and ovarian cancer risks using information on personal risk factors, cancer family history, genetic testing for high- and moderate-risk genes, polygenic risk scores and mammographic density. Current risk prediction models for breast/ovarian cancers have several limitations. They often do not account for risk-reducing interventions and/or the age those interventions were performed (e.g., RRSO). They fail to incorporate the whole family history of disease, in particular the exact age at onset of the cancers in the relatives. As clinicians are moving to a more personalized management of women with high-risk pathogenic variants, it is important to incorporate those information into risk prediction models. Our goal is to propose a more comprehensive prediction model for the personalized clinical management of women with PVs in {\it BRCA1}. This model is able to estimate the risk of BC accounting for the full history of this cancer within the family, the personal and family history of RRSO as well as the exact age of this intervention among family members. This work is based on the analysis of data from the Breast Cancer Family Registry (BCFR). }

\noindent From a statistical point of view, the response variable in this dataset is the age at onset of BC. This variable is right-censored as some women had not experienced the event of interest by the last follow-up. Two explanatory variables are considered in this work. The first is the age at which RRSO was performed. The second covariate is a woman's genotype corresponding to the P/LPV carrier status. In this work, we consider a Cox model to investigate the impact of these explanatory variables on the age at onset of BC.
\noindent \textcolor{black}{Several methodological issues complicate the statistical analysis of the BCFR data.} First, observations are clustered into families, which induces a correlation between the outcomes of individuals belonging to the same family. We model these dependencies using a Gaussian copula indexed by a matrix whose entries are expressed in terms of kinship coefficients and an heritability parameter. Second, the age at which RRSO was performed is a time-varying covariate. Choi et al. \cite{choi2021competing} considered a parametric approach to model the impact of this covariate and found out that the results are sensitive to the specification of the parametric form. In this work, we model the impact of RRSO on BC through B-splines. These B-splines are expressed in terms of the lag between the current age and the age at which RRSO was performed. Finally, families are ascertained via a proband who is affected with cancer by the time of recruitment. Therefore, one has to apply a correction for ascertainement in order to obtain valid inference about the marginal and association parameters of our model. In this work, we develop an iterative algorithm to estimate the parameters of the model in the presence of a selection bias. The proposed procedure extends the two stage procedure \citep{andersen2005two} and iterates between the maximizations of an univariate and a bivariate log-likelihood functions until convergence. 

\noindent The remainder of the paper is organized as follows. In Section 2, we present the data at hand. In section 3, we provide the details of our model. Section 4 presents the proposed iterative algorithm to estimate the parameters of the model. The proposed method is applied to the data at hand and the results are reported in Section 5. Section 6 reports an extensive simulation study to investigate the finite sample properties of the proposal. Conclusions and final thoughts are given in Section 7.


\section{Data}
\noindent \textcolor{black}{We consider data from BCFR, which was established in 1995 with participants from USA, Australia and Canada \citep{john2004}. Families with {\it BRCA1} mutations and multiple cases of breast and/or ovarian cancers were enrolled. The data were mostly collected between 1996 and 2000, with few families included afterwards. This is a family-based cohort over-sampled for increased BC familial risk. Each sampled family includes a proband, i.e. the initial member of the family to be identified, as well as first and second degree relatives. During follow-up, extensive information was collected on ages at the breast or ovarian cancer diagnosis since enrollment in the BCFR, surgeries, and mammographic screening collected from baseline and follow-up questionnaires as well as pathogenic variant status in {\it BRCA1} and {\it BRCA2} genes \citep{terry2016}. We restricted our data analyses to families that were known to carry {\it BRCA1} pathogenic variants. We assumed that all family members entered the study at the age of 18 and were followed until the onset of breast cancer, or the age at last follow-up. All centers in these consortia obtained written informed consent from study participants and local ethical review committees approved protocols.}

\section{Notation and model}

\noindent Consider $I$ independent families and for $i=1,\cdots,I$, we index the $n_i+1$ members of family $i$ by the subscript $j \in \{0,1\cdots,n_i\}$. We let the subscript $j=0$ correspond to the proband. For $j=0,\cdots,n_i$, let $T_{ij}$ be the age at BC onset and $X_{ij}=\{X_{ij1},X_{ij2}\}$ be a pair of covariates for individual $j$ in family $i$. In our dataset, $X_{ij1}$ is the age at which RRSO was performed, with $X_{ij1}=\infty$ if no oophorectomy was performed, and $X_{ij2}$ is an individual's pathogenic variant status (carrier=1, noncarrier=0). The total size of the dataset is $N=I+\sum\limits_{i=1}^I n_i$.

\subsection{Marginal distribution}

\noindent We consider the following proportional hazard\textcolor{black}{s} Cox model
\begin{align}
    \lambda(t|X_{ij})=\lambda_0(t)\exp\left\{g(t,X_{ij1}) +\beta X_{ij2}\right\}, \label{model1}
\end{align}
where $\lambda_0(t)=\alpha \lambda t^{\alpha-1}$ is the baseline hazard function specified according to the Weibull distribution with parameters $\alpha$ and $\lambda$. In our formulation, we assume that $$g(t,X_{ij1})=r(t-X_{ij1}) \mathbf{1}_{\{t>X_{ij1}\}},$$ where $r:\mathbb{R}^{+} \rightarrow \mathbb{R}$ is a completely unspecified function such that $r(0)=0$. Our model assumes that the effect of oophoractomy depends only on the gap time between the present time and the age at oophoractomy. The cumulative hazard, the survival and the density functions are given by $\Lambda(t|X_{ij})=\int_0^t\lambda(u|X_{ij}) du$, $S(t|X_{ij})=\exp\{- \Lambda(t|X_{ij})\}$ and $ f(t|X_{ij})=\lambda(t|X_{ij}) \times S(t|X_{ij})$, respectively.

\subsection{Familial relatedness and dependence structure}
\noindent We model the relatedness in family $i$ by assuming that, given the covariates, the joint distribution of $(T_{i0},T_{i1},\cdots,T_{in_i})$ follows a Gaussian copula $D$ indexed by a correlation matrix $V^{(i)}$. This is equivalent to specifying that $(Z_{i0},Z_{i1},\cdots,Z_{in_i})$ follows a multivariate normal distribution with zero mean and covariance matrix $V^{(i)}$, where $Z_{ij}=\Phi^{-1}\left[S(T_{ij}|X_{ij})\right], j=0,\cdots,n_i$ are the inverse normal scores and $\Phi^{(-1)}$ is the quantile function of the standard normal distribution. The off-diagonal entry $(j,k)$ of $V^{(i)}$ captures the polygenic heritability parameter between individuals $j$ and $k$ in family $i$. This term is assumed to be equal to $$V_{jk}^{(i)}=hR_{jk}^{(i)},$$ where $h \in [0,1]$ is a heritability parameter, equal to the percentage of variance explained by the familial relatedness, $R_{jk}^{(i)} = 2 \times \phi_{jk}^{(i)}$ the relatedness coefficient and $\phi_{jk}^{(i)}$ the kinship coefficient between individuals $j$ and $k$ in family $i$. The diagonal entries of $V^{(i)}$ are equal to 1. Using matrix notation, one has
$$V^{(i)} = hR^{(i)}+(1-h)I_{n_i+1},$$
where $I_{n_i+1}$ is the identity matrix of size $n_i+1$ and $R^{(i)}$ is a square matrix with  diagonal elements equal to 1 and off-diagonal elements equal to $R_{jk}^{(i)}$. For simplicity, we omit the dimension and the parameters of the copula $D$. Under the considered model, one has
$$P(T_{i0}>t_{i0},\cdots,T_{in_i}>t_{in_i}|X_{i0},\cdots,X_{in_i})=D\left\{S(t_{i0}|X_{i0}),\cdots,S(t_{in_i}|X_{in_i})\right\}.$$

\subsection{Useful properties of the model}

\noindent In this section, we present the expressions of the conditional distributions for the times $T_{ij}$ and the pairs of times at onset of BC $(T_{ij},T_{ik})$ given the information that we have on the affected proband. These expressions are required later for the construction of the likelihood functions.

\noindent For $i=1,\cdots,I$, $j=1,\cdots,n_i$, one has
$$P(T_{i0}>t_{i0},T_{ij}>t_{ij}|X_{i0},X_{ij}) = D\{S(t_{i0}|X_{i0}),S(t_{ij}|X_{ij})\}.$$
Therefore the conditional survival and density functions of $T_{ij}$ given $T_{i0}=t_{i0}$ are respectively equal to
$$P(T_{ij}>t_{ij}|T_{i0}=t_{i0},X_{ij},X_{i0}) = D_{10}\{S(t_{i0}|X_{i0}),S(t_{ij}|X_{ij})\}$$
and
$$-\frac{\partial}{\partial t}P(T_{ij}>t|T_{i0}=t_{i0},X_{ij},X_{i0})|_{t=t_{ij}} = D_{11}\{S(t_{i0}|X_{i0}),S(t_{ij}|X_{ij})\} f(t_{ij}|X_{ij}),$$
where $D_{10}(u,v)=\partial D(u,v)/\partial u$ and $D_{11}(u,v)=\partial^2 D(u,v)/\partial u \partial v$.

\noindent Moreover, for $i=1,\cdots,I$, $1 \le j < k \le n_i$, one has
$$P(T_{i0}>t_{i0},T_{ij}>t_{ij}, T_{ik}>t_{ik}|X_{i0},X_{ij},X_{ik}) = D\{S(t_{i0}|X_{i0}),S(t_{ij}|X_{ij}),S(t_{ik}|X_{ik})\}.$$

\noindent Therefore, the joint distribution of $(T_{ij},T_{ik})$ given $T_{i0}=t_{i0}$ is characterized by
$$P(T_{ij}>t_{ij},T_{ik}>t_{ik}|T_{i0}=t_{i0}) = D_{100}\{S(t_{i0}|X_{i0}),S(t_{ij}|X_{ij}),S(t_{ik}|X_{ik})\},$$
$$ -\frac{\partial}{\partial t} P(T_{ij}>t,T_{ik}>t_{ik}|T_{i0}=t_{i0}))|_{t=t_{ij}} = D_{110}\{S(t_{i0}|X_{i0}),S(t_{ij}|X_{ij}),S(t_{ik}|X_{ik})\} f(t_{ij}|X_{ij}),$$
and
$$\frac{\partial^2}{\partial t \partial s} P(T_{ij}>t,T_{ik}>s|T_{i0}=t_{i0})|_{t=t_{ij},s=t_{ik}} = D_{111}\{S(t_{i0}|X_{i0}),S(t_{ij}|X_{ij}),S(t_{ik}|X_{ik})\} f(t_{ij}|X_{ij})f(t_{ik}|X_{ik}),$$
where $D_{100}(u,v)=\partial D(u,v,w)/\partial u$, $D_{110}(u,v)=\partial^2 D(u,v,w)/\partial u \partial v$, and $D_{111}(u,v,w)=\partial^3 D(u,v,w)/\partial u \partial v \partial w$.

\section{Inference procedures}

\subsection{Observed data}

\noindent For $i=1,\cdots,I$, let $a_i$ be the age at examination of the proband in family $i$. According to the data collection protocol, family $i$ is included in the data set if and only if the proband has an observed BC prior to the examination age. Therefore, the inclusion condition for family $i$ is $\{T_{i0} < a_i\}$. For the nonprobands, the exact value of $T_{ij}$ is not always observed due to right-censoring. For $i=1,\cdots,I$; $j=1,\cdots,n_i$, we only observe $Y_{ij}=\mbox{min}(T_{ij},C_{ij})$ and $\delta_{ij}=\mathbf{1}_{\{T_{ij}<C_{ij}\}}$, where $C_{ij}$ is the censoring time for individual $j$ in family $i$, assumed  independent of $T_{ij}$. Therefore, the observed data is $O=\{O_1,\cdots,O_I\}$, where $O_i=\{a_i,(Y_{i0},\tilde X_{i0},\delta_{i0}),\cdots,(Y_{in_i},\tilde X_{in_i},\delta_{in_i})\},$
$\tilde X_{ij}=(\tilde X_{ij1},X_{ij2})$ and $\tilde X_{ij1}=\mbox{min}(X_{ij1},C_{ij})$.

\subsection{Parameters estimation}

\noindent In this work, we use B-splines to express the unknown function $r$. Accordingly, one writes $r(u)=\sum\limits_{l=1}^L b_l B_l(u)$, where $\{B_1 \textcolor{black}{(u)},\cdots,B_L\textcolor{black}{(u)}\}$ is a B-splines basis \textcolor{black}{function} of dimension $L$ such that $B_l(0)=B_l'(0)=B_l''(0)=0$ for $l=1,\cdots,L$.  Therefore, the set of parameters to be estimated from the data is $\Theta=\{\theta,h\}$, where $\theta=\{\alpha,\lambda,\beta,b_1,\cdots,b_L\}$. Note that the set of parameters $\theta$ characterizes the marginal distribution of the age at onset of BC whereas $h$ measures the association between these ages within a family.

\noindent We use the likelihood principle to estimate $\Theta$. The data collection protocol induces a selection bias and therefore an ascertainment correction is required. To this end, one has to maximize the conditional full likelihood where the contribution of each family is corrected for its probability of being ascertained. Writing and maximizing the resulting full conditional log-likelihood function yields very unstable results with the data at hand.

\noindent In this work, we consider two conditional log-likelihood functions $l^{(1)}(\theta|h,O)$ and $l^{(2)}(h|\theta,O)$, which we denote by univariate and bivariate conditional log-likelihood functions, respectively. The spirit of these functions comes from the two-stage procedure, where the marginal parameters are estimated by maximizing a univariate log-likelihood function first and the association parameter is obtained by maximizing a bivariate log-likelihood function, given the estimates of the marginal parameters, afterwards \cite{andersen2005two}. In our context, the univariate conditional likelihood function depends, not only on $\theta$, but also on $h$ due to the ascertainment correction. Therefore, it is impossible to apply the two-stage procedure directly. We generalize the latter procedure by proposing an algorithm that iterates between the maximization of $l^{(1)}(\theta|h,O)$ and $l^{(2)}(h|\theta,O)$, until convergence. 

\subsubsection{Univariate log-likelihood function}

\noindent The univariate conditional log-likelihood function is the sum of log-contributions from each of the $N$ individuals. It can written as $$l^{(1)}(\theta|h,O)=\sum\limits_{i=1}^I l^{(1)}_{i}(\theta|h,O_i) = \sum\limits_{i=1}^I  \sum\limits_{j=0}^{n_i} l^{(1)}_{ij}(\theta|h,O_i) = \sum\limits_{i=1}^I  \sum\limits_{j=0}^{n_i} \log\left\{L^{(1)}_{ij}(\theta|h,O_i)\right\} .$$

\noindent The contribution of the proband of family $i$ to the univariate conditional likelihood function is 
$$L^{(1)}_{i0}(\theta|h,O_i) = f(y_{i0}|T_{i0}<a_i,X_{i0})=\frac{f(y_{i0}|X_{i0})}{1-S(a_i|X_{i0})}.$$

\noindent On the other hand, for nonproband $j$, $j=1,\cdots,n_i$, the expression of $L^{(1)}_{ij}(\theta|h,O_i)$ depends on $\delta_{ij}$. 

\noindent If $\delta_{ij}=0$, the event is censored and 
$$L^{(1)}_{ij}(\theta|h,O_i)=P(T_{ij}>y_{ij}|T_{i0}=y_{i0},X_{ij},X_{i0})=D_{10}\{S(y_{i0}|X_{i0}),S(y_{ij}|X_{ij})\}.$$

\noindent If $\delta_{ij}=1$, the event is observed and 
\begin{eqnarray*}
L^{(1)}_{ij}(\theta|h,O_i)&=&-\left\{\frac{\partial}{\partial y} P(T_{ij}>y|T_{i0}=y_{i0},X_{ij},X_{i0})\right\}_{|y=y_{ij}} \\
&=&D_{11}\left\{S(y_{i0}|X_{i0}),S(y_{ij}|X_{ij})\right\}f(y_{ij}|X_{ij}).
\end{eqnarray*}
 
Therefore, one has

\begin{eqnarray*}
l_{i}^{\textcolor{black}{(1)}}(\theta|h,O_i) &=& \log\left\{f(y_{i0}|X_{i0})\right\} -  \log\left\{1-S(a_i|X_{i0})\right\} \\
&+& \sum_{j=1}^{n_i} \delta_{ij} \log\left[D_{11}\left\{S(y_{i0}|X_{i0}),S(y_{ij}|X_{ij})\right\} f(y_{ij}|X_{ij})\right] \\
&+& \sum_{j=1}^{n_i} (1-\delta_{ij}) \log\left[D_{10}\{S(y_{i0}|X_{i0}),S(y_{ij}|X_{ij})\}\right].
\end{eqnarray*}

\subsubsection{Bivariate log-likelihood function}

\noindent The bivariate conditional log-likelihood function is the sum of log-\textcolor{black}{likelihood} contributions from each of the $\sum\limits_{i=1}^I {{n_i}\choose{2}}$ pairs of nonproband individuals. Therefore, one has $$l^{(2)}(h|\theta,O)=\sum\limits_{i=1}^I l^{(2)}_{i}(h|\theta,O_i) = \sum\limits_{i=1}^I \sum\limits_{1\le j < k  \le n_i} l^{(2)}_{ijk}(h|\theta,O_i).$$

\noindent In the Appendix \textcolor{black}{A}, we show that $l^{(2)}(h|\theta,O)$ is equal to
$$ \sum\limits_{i=1}^I \sum\limits_{1\le j < k  \le n_i} \log\left[ D_{1\delta_{\textcolor{black}{ij}}\delta_{\textcolor{black}{ik}}}\left\{S(t_{i0}|X_{i0}),S(t_{ij}|X_{ij}),S(t_{ik}|X_{ik})\right\}\right]$$
up to an additive constant that does not depend on $h$.

\subsubsection{Estimation procedure}

\noindent We propose to estimate $\Theta=(\theta^\top,h)^\top$ using the following iterative algorithm:
\begin{enumerate}
\item Obtain an initial estimate of $\theta$ by maximizing $l^{(1)}(\theta|h=0,O)$.
\item Estimate $h$ by maximizing $l^{(2)}(h|\theta,O)$.
\item Estimate $\theta$ by maximizing $l^{(1)}(\theta|h,O)$.
\item Iterate between steps 2. and 3. until convergence.
\end{enumerate}

\noindent At convergence, we obtain $\hat\Theta=(\hat\theta^\top,\hat h)^\top$. 

\subsubsection{Penalized log-likelihood}

\noindent As will be seen in the application to {\it BRCA1} families data, the obtained estimates of the B-splines coefficients $b_1,\cdots,b_L$ have high variances. This is due to the repartition of the ages at onset of the observed events within the space definition of the B-splines. One solution to this problem is to consider penalized log-likelihood, which corresponds to adding  a penalty term $-\gamma \sum_{j=3}^{L} \left(\Delta^2 b_j\right)^2$ to $l^{(1)}(\theta|h,O)$ with $\Delta^2 b_j = b_j - 2b_{j-1} + b_{j-2}$, where $\gamma$ is a smoothing penalty parameter \citep{eilers2021practical}.

\subsubsection{Variance-covariance matrix estimation}

\noindent The variance--covariance matrix of the joint estimator
$\hat{\Theta} = (\hat{\theta}^\top, \hat{h})^\top$
is obtained using asymptotic M-estimation theory via a sandwich (robust) estimator.
The estimator is defined as the solution of the stacked score equation
$M_I(\Theta) = 0$, where
\[
M_I(\Theta) = \frac{1}{I} \sum_{i=1}^I m_i(\Theta)
\]
is the empirical estimating function and $m_i(\Theta)$ denotes the contribution of
independent unit $i$ (e.g., family $i$). The stacked score vector is defined as
\[
m_i(\Theta) =
\begin{pmatrix}
\frac{\partial l_i^{(1)}(\theta \mid h, O_i)}{\partial \theta} \\[1.2em]
\frac{\partial l_i^{(2)}(h \mid \theta, O_i)}{\partial h}
\end{pmatrix},
\]
where $l^{(1)}$ and $l^{(2)}$ denote the univariate and bivariate log-likelihood
contributions, respectively.

\noindent Under standard regularity conditions, a first-order Taylor expansion of
$M_I(\hat{\Theta})$ around the true parameter value $\Theta_0$ yields the asymptotic
linear representation
\[
\sqrt{I}\,(\hat{\Theta} - \Theta_0)
=
A^{-1}
\left(
\frac{1}{\sqrt{I}} \sum_{i=1}^I m_i(\Theta_0)
\right)
+ o_p(1),
\]
where

\textcolor{black}{
\[
A
=
\mathbb{E}
\left[
-\frac{\partial M_I(\Theta)}{\partial \Theta^\top}
\right]_{\Theta=\Theta_0}.
\]
}

\noindent By the central limit theorem,
\[
\sqrt{I}\,(\hat{\Theta} - \Theta_0)
\;\xrightarrow{\mathcal{L}}\;
\mathcal{N}\!\left(0,\, A^{-1} B A^{-1}\right),
\]
where

\textcolor{black}{
\[
B
=
\mathbb{E}
\left[
M_I(\Theta_0) M_I(\Theta_0)^\top
\right].
\]
}

\noindent In practice, $A$ and $B$ are replaced by their empirical counterparts evaluated
at $\hat{\Theta}$,
\[
\hat{A}
=
\textcolor{red}{-}
\frac{1}{I}
\sum_{i=1}^I
\left.
\frac{\partial m_i(\Theta)}{\partial \Theta^\top}
\right|_{\Theta=\hat{\Theta}},
\qquad
\hat{B}
=
\frac{1}{I}
\sum_{i=1}^I
m_i(\hat{\Theta}) m_i(\hat{\Theta})^\top .
\]

\noindent The resulting sandwich estimator of the asymptotic covariance matrix of
$\hat{\Theta}$ is
\[
\widehat{\mathrm{Var}}(\hat{\Theta})
=
\frac{1}{I}
\hat{A}^{-1}
\hat{B}
(\hat{A}^{-1})^\top.
\]

\section{Application to {\it BRCA1} families from BCFR}

\subsection{Analysis}

\noindent Our data consists of 346 families comprising 1706 individuals with a follow-up age ranging from 18.14 to 102.5 years of age. The numbers of individuals per family are given in Appendix A. 
The dataset includes all \textit{BRCA1} families for which the proband had been diagnosed with BC prior to study entry.
In total, 683 BCs (min age: 21.84, mean age: 43.88, max age: 85.10) and 106 oophorectomy events (min age: 24.50, mean age: 45.03, max age: 75.5) were observed. Finally, 502 women among the 1706 participants were carriers of a pathogenic variant. We applied the proposed method to the family data with $L=10$ B-splines and various values of $\gamma$. The results are reported in Table 1, below.

\begin{table}[H]
\centering
\footnotesize
\begin{tabular}{|c|cc|cc|cc|cc|cc|}
\hline
 & \multicolumn{2}{c|}{$\gamma=0$} 
 & \multicolumn{2}{c|}{$\gamma=0.05$} 
 & \multicolumn{2}{c|}{$\gamma=0.1$} 
 & \multicolumn{2}{c|}{$\gamma=0.5$} 
 & \multicolumn{2}{c|}{$\gamma=1$} \\
\hline
Parameters 
& Estimate & s.e. 
& Estimate & s.e. 
& Estimate & s.e. 
& Estimate & s.e. 
& Estimate & s.e. \\
\hline
$\log(\alpha)$ 
& 1.077 & 0.042 & 1.079 & 0.042 & 1.078 & 0.042 & 1.078 & 0.042 & 1.077 & 0.042 \\

$\log(\lambda)$ 
& -13.664 & 0.746 & -13.696 & 0.741 & -13.686 & 0.739 & -13.690 & 0.738 & -13.682 & 0.735 \\

$\beta_{\text{gene}}$ 
& 1.939 & 0.214 & 1.941 & 0.214 & 1.939 & 0.214 & 1.937 & 0.213 & 1.937 & 0.226 \\

$b_1$ 
& -1.144 & 1.562 & -1.000 & 1.340 & -1.108 & 1.375 & -1.273 & 1.205 & -1.240 & 1.121 \\

$b_2$ 
& -0.363 & 2.353 & -0.908 & 0.895 & -0.876 & 0.785 & -0.705 & 0.667 & -0.644 & 3.430 \\

$b_3$ 
& -1.023 & 1.809 & -0.367 & 0.732 & -0.220 & 0.630 & -0.036 & 0.451 & -0.025 & 5.414 \\

$b_4$ 
& 1.324 & 0.808 & 0.981 & 0.540 & 0.842 & 0.532 & 0.555 & 0.466 & 0.473 & 2.091 \\

$b_5$ 
& -0.130 & 1.266 & 0.028 & 1.255 & 0.148 & 0.550 & 0.399 & 0.411 & 0.466 & 0.492 \\

$b_6$ 
& -0.261 & 1.503 & 0.233 & 0.727 & 0.307 & 0.560 & 0.455 & 0.395 & 0.492 & 0.398 \\

$b_7$ 
& 1.796 & 1.084 & 1.000 & 0.547 & 0.836 & 0.520 & 0.536 & 0.455 & 0.438 & 0.541 \\

$b_8$ 
& -3.230 & 3.588 & -0.952 & 0.686 & -0.627 & 0.573 & -0.261 & 0.485 & -0.206 & 0.814 \\

$b_9$ 
& 0.923 & 5.688 & -1.243 & 1.453 & -1.273 & 1.380 & -1.088 & 0.973 & -0.942 & 1.080 \\

$b_{10}$ 
& -0.988 & 6.242 & -0.663 & 3.314 & -1.391 & 3.066 & -1.825 & 1.808 & -1.651 & 1.634 \\

$h$ 
& 0.527 & 0.263 & 0.544 & 0.257 & 0.547 & 0.257 & 0.552 & 0.256 & 0.554 & 0.256 \\

\hline

\end{tabular}
\caption{Estimates and standard errors for the model parameters under different values of the smoothing parameter $\gamma$. The parameters $b_1, b_2,\dots , b_{10}$ denote the coefficients of the B-spline basis functions of the B-spline model ($L=10$, order $=4$)}.
\label{tab:gamma10}
\end{table}

\noindent As we can see from this Table, the estimates of $\log(\alpha)$, $\log(\lambda)$, $\beta_{gene}$ and $h$ and their standard errors remain stable  across various values of $\gamma$. 

\noindent When $\gamma=0$, the obtained estimates of the B-spline coefficients $b_1,\cdots,b_{10}$ are highly variables. The illustration of this phenomena is more clear in Figures 1 and 2 below. This is due to the fact that we only have 26 observations of oopheractomy events to estimate these parameters. This prompted us to consider a penalized likelihood approach \citep{murphy2012machine}. 

\noindent A key step of the penalized approach is the selection of the smothing parameter $\gamma$. In the litterature, a common method for choosing $\gamma$ consists on selecting the value that minimizes some estimable criterion function such as the integrated Brier score. Typically, the generalized cross validation procedure is used to provide a numerical estimate of the criteria. Unfortunetly, such an approach is difficult to adapt to our context as no simple criterion function seem to be available in the presence of familial dependencies and ascertainement bias. 

\noindent In the simulations section, we show that the considered iterative algorithm yields consistent estimates with $\gamma=0$ if enough observations are available. Moreover, we show that the penalized likelihood approach introduces a bias that increases with the smoothing parameter $\gamma$. Therefore, in this application, we apply a rule of thumb and select the smallest value of $\gamma$ that stabilises the variances of the estimates of $b_1,\cdots,b_L$. The value $\gamma=0.05$ works well and yields raisonable results.

\noindent In Figure 1, we present the estimates of $S(t|X_{ij1}=x_{ij1};X_{ij2}=1)$ as a function of $t$ obtained with $\gamma \in \{0, 0.05, 0.5, 1\}$. The 4 curves in each plot correspond to $x_{ij1}$ equal to $30$, $35$, $40$ , $45$ and $\infty$ years, respectively. In Figure 2, we report the estimates of $r(u)$ as a function of $u$, with the $95\%$ confidence bands, obtained with $\gamma \in \{0, 0.05, 0.5, 1\}$.


\begin{figure}[H]
   \centering
 \includegraphics[scale=0.9]{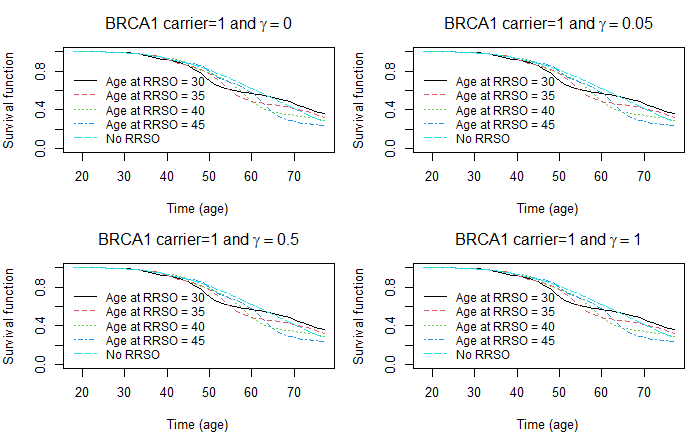}
 \caption{Estimate of $S(t|X_{ij})$, as a function of $t$, with $X_{ij1}$ equal to $30$, $35$, $40$, $45$ and $\infty$, and $X_{ij2}=1$ under different values of the smoothing parameter $\gamma$.} 
\end{figure}


\begin{figure}[H]
   \centering
 \includegraphics[scale=0.8]{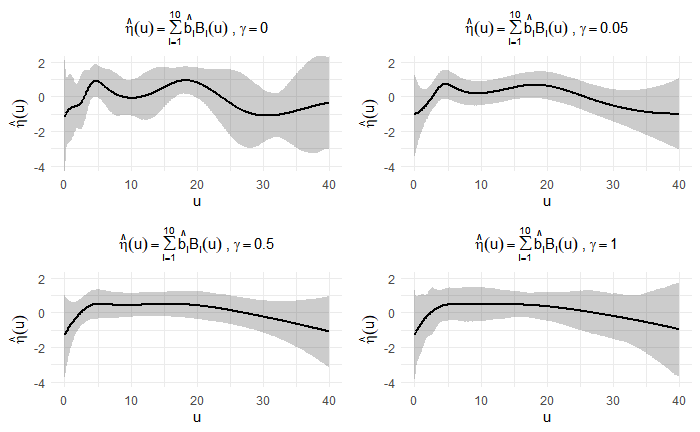}
 \caption{Estimates of $r(u)$, as a function of $u$, along with the $95\%$ confidance bands, obtained with $\gamma \in \{0,0.05,0.5,1\}$.} 
\end{figure}

\noindent These figures highlight the complex relationship between RRSO and BC risk. \textcolor{black}{Having an oophorectomy tends to slightly increase BC risk shortly after RRSO but then this risk decreases smoothly over time.} However, \textcolor{black}{it is difficult to investigate this relationship further} because only 26 oophorectomy were observed for 1706 individuals.

\subsection{Risk functions}

\noindent \textcolor{black}{Conducting familial studies allows us to incorporate BC familial history as well as RRSO family hisotry into an individual's risk estimates.} Typically, risk functions evaluated at age $t$ are defined as conditional cumulative distribution functions given familial history and individual information collected up to age $t$. Using the model considered in this work, one may derive expressions for these risks functions.
\noindent Once the parameters $\Theta = (\theta^{\top}, h)^{\top}$ are estimated from the data, it is possible to estimate and graphically represent the risk functions using the plug-in principle. It is possible then to obtain confidence bands using the delta method.

\noindent  In the following sections, we consider three situations \textcolor{black}{that might be of clinical interest} and for each of them, we define a risk function and present the obtained BC estimates.

\subsubsection{Kinship-dependent BC risk prediction}

\noindent Consider for instance a pair of individuals $j$ and $k$ from family $i$ with a kinship coefficient $\phi^{(i)}_{jk}$. Assume that we want to estimate BC risk of individual $j$ given the cancer \textcolor{black}{status} of individual $k$. We assume that individual $j$ had been followed up until the age of $t_{ij}=35$ years. She is BC free and had RRSO at age $X_{ij1}=32$ years. We also assume that individual $k$ had cancer at the age of $T_{ik}=45$ years and never had RRSO, i.e. $X_{ik1}=\infty$. 
For $t>35$, the risk function of individual $j$ given \textcolor{black}{the BC status} of $k$ is given by 
\begin{eqnarray*}
risk(t) &=& P(T_{ij}<t | T_{ij}>35, T_{ik}=45, X_{ij1}=32, X_{ij2}, X_{ik1}=\infty, X_{ik2}) \\
&=& 1-P(T_{ij}>t | T_{ij}>35, T_{ik}=45, X_{ij1}=32, X_{ij2}, X_{ik1}=\infty, X_{ik2}) \\
&=& 1-\frac{D_{01}\{S(t|X_{ij1}=32, X_{ij2}),S(45|X_{ik1}=\infty, X_{ik2})\}}{D_{01}\{S(35|X_{ij1}=32, X_{ij2}),S(45|X_{ik1}=\infty, X_{ik2})\}}
\end{eqnarray*}

\noindent In Figure \textcolor{black}{3}, we present the estimates of $risk(t)$  
as function of $t$ when $X_{ij2}=X_{ik2}=0$ \textcolor{black}{(i.e., individuals $j$ and $k$ are noncarriers)}. The four curves in this Figure correspond to a kinship $\phi^{(i)}_{jk}$ equal to $0$ (unrelated), $0.0625$ (first cousins), $0.125$ (aunt) and $0.25$ (siblings or mother/daughter), respectively. Figures corresponding to other values of $(X_{ij2},X_{ik2})$ are given in Appendix D.

\begin{figure}[H]
   \centering
 \includegraphics[scale=0.85]{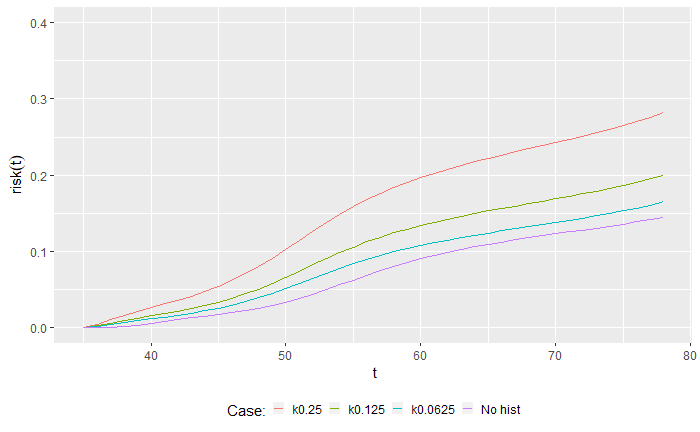}
 \caption{Estimate of $risk(t)$ as a function of $t$ when $X_{ij2}=32$, $X_{ik2}=\infty$, and $X_{ij2}=X_{ik2}=0$.}
\end{figure}

\noindent These figures clearly illustrae that an individual with a relative who had been diagnosed with BC has an increased risk of developing the disease. Moreover, this risk increases with the kinship coefficient. A closer relative with a positive BC history implies an augmented risk.

\subsubsection{Age-specific BC risk prediction}

\noindent Consider now a pair of individuals $j$ and $k$ from family $i$ such as $\phi^{(i)}_{jk}=0.25$. We assume that individual $j$ had been followed up until the age of $t_{ij}=35$ years. She is cancer free and had RRSO at age $X_{ij1}=32$ years. We also assume that individual $k$ had BC at the age of $T_{ik}=s$ years and never had oophoractomy, i.e. $X_{ik1}=\infty$. 
\noindent For $t>35$, the $\omega$-year risk function of individual $j$ given the cancer history of individual $k$ is given by 
\begin{eqnarray*}
risk_\omega(s) &=& P(T_{ij}<35+\omega | T_{ij}>35, T_{ik}=s, X_{ij1}=32, X_{ij2}, X_{ik1}=\infty, X_{ik2}) \\
&=&1- P(T_{ij}>35+\omega | T_{ij}>35, T_{ik}=s, X_{ij1}=32, X_{ij2}, X_{ik1}=\infty, X_{ik2}) \\
&=&1- \frac{D_{01}\{S(35+\omega|X_{ij1}=32, X_{ij2}),S(s|X_{ik1}=\infty, X_{ik2})\}}{D_{01}\{S(35|X_{ij1}=32, X_{ij2}),S(s|X_{ik1}=\infty, X_{ik2})\}}.
\end{eqnarray*}

\noindent In Figure \textcolor{black}{4}, we present estimates of $risk_\omega(s)$ and $-\log[risk_\omega(s)]$ as functions of $s$ when $X_{ij2}=X_{ik2}=0$. The three curves in this Figure correspond to $\omega$ equal to 5, 10 and 15 years, respectively. Figures corresponding to other values of $(X_{ij2},X_{ik2})$ are given in Appendix E.

\begin{figure}[H]
   \centering
 \includegraphics[scale=0.85]{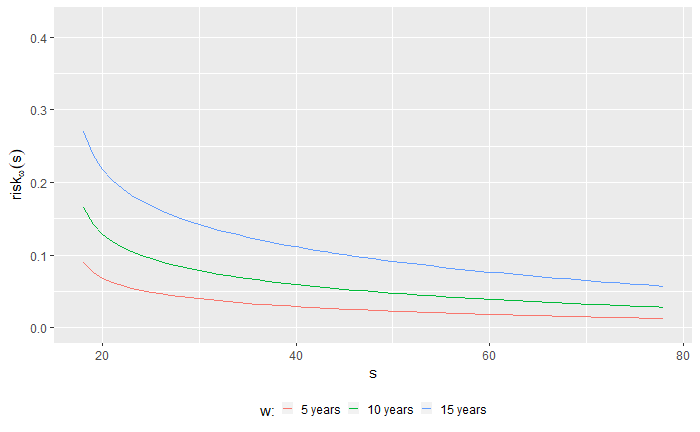}
 \caption{Estimate of $risk_\omega(s)$ as a function of $s$ when $X_{ij1}=32$, $X_{ik1}=\infty$, and $X_{ij2}=X_{ik2}=0$. The three curves correspond to $\omega=5, 10$ and $15$ years, respectively.}
\end{figure}
\noindent This Figure illustrates how the 5, 10 and 15-year risks for individual $j$ evolve as a function of the age-at-onset of cancer for individual $k$. The younger individual $k$ had BC, the higher is the BC risk for individual $j$.

\subsubsection{\textcolor{black}{BC risk prediction integrating paternal vs. maternal side of BC family history}}

\noindent Consider for instance a family $i$ with three members, a daughter $j$, her mother $k$ and her aunt $l$. We assume that individual $j$ had been followed up until the age of $t_{ij}=35$ years. She is cancer free and had RRSO at age $X_{ij1}=32$ years. We also assume that individuals $k$ and $l$ were both diagnosed with BC at the age of 45 years and never had oophorectomy. For $t>35$, the risk function of individual $j$ given the cancer history of $k$ and $l$ is given by 
\begin{eqnarray*}
risk(t) &=& P(T_{ij}<t | T_{ij}>35, T_{ik}=45, T_{il}=45, X_{ij1}=32, X_{ij2}, X_{ik1}=\infty, X_{ik2}, X_{il1}=\infty, X_{il2}) \\
 &=&1- P(T_{ij}>t | T_{ij}>35, T_{ik}=45, T_{il}=45, X_{ij1}=32, X_{ij2}, X_{ik1}=\infty, X_{ik2}, X_{il1}=\infty, X_{il2}) \\
&=& 1-\frac{D_{01}\{S(t|X_{ij1}=32, X_{ij2}),S(45| X_{ik1}=\infty, \textcolor{black}{X_{ik2}}),S(45|X_{il1}=\infty, X_{il2})\}}{D_{01}\{S(35|X_{ij1}=32, X_{ij2}),S(45|X_{ik1}=\infty, X_{ik2}),S(45|X_{il1}=\infty, X_{il2})\}}.
\end{eqnarray*}

\noindent Let us now distinguish two cases depending on whether the aunt is maternal or paternal. If Aunt $l$ is maternal, then individuals $k$ and $l$ are sisters and the kinship matrix of family $i$ is 
$$\phi^{(i)} = \begin{pmatrix}  0.5 & 0.25 & 0.125 \\ 0.25 & 0.5 & 0.25 \\ 0.125 & 0.25 & 0.5 \end{pmatrix}.$$ 
On the other hand, if aunt $l$ is paternal, then individuals $k$ and $l$ are unrelated and the kinship matrix of family $i$ is
$$\phi^{(i)} = \begin{pmatrix}  0.5 & 0.25 & 0.125 \\ 0.25 & 0.5 & 0 \\ 0.125 & 0 & 0.5 \end{pmatrix}.$$

\noindent In Figure 5, we compare the estimates of $risk(t)$
, as functions of $t$, obtained with these kinship matrices when $X_{ij2}=X_{ik2}=X_{il2}=0$. Figures corresponding to other values of $(X_{ij2},X_{ik2},X_{il2})$ are given in  Appendix F.

\begin{figure}[H]
   \centering
 \includegraphics[scale=0.85]{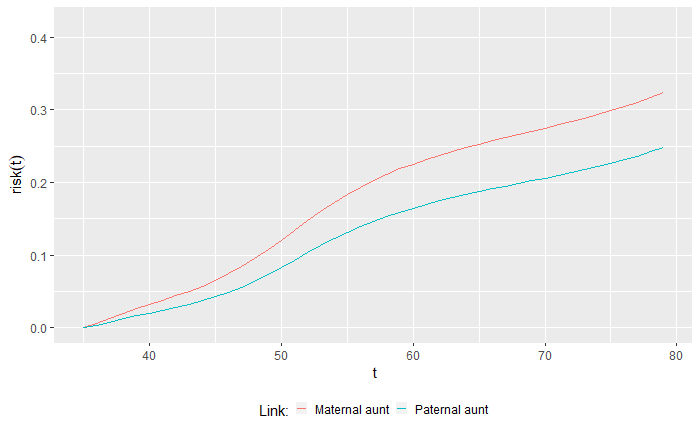}
 \caption{Estimate of $risk(t)$ as a function of $t$ when $X_{ij1}=32$, $X_{ik1}=X_{il1}=\infty$, and $X_{ij2}=X_{ik2}=X_{il2}=0$. The orange curves refer to estimates when the aunt is maternal whereas the blue ones are for the case of a paternal aunt.}
\end{figure}

\noindent All these figures suggest an increased BC risk when the aunt is maternal. 

\section{Simulations}

\noindent We conducted a simulation study to assess the empirical properties of the proposed iterative algorithm to estimate the parameters of the consider model. We considered a setting with $I$ families each having an identical structure. Each family has 3 pairs of siblings and each pair of individuals who are not sibling are assumed to be cousins. Therefore, we have the following kinship matrix: 
$$
\phi=
\begin{pmatrix}
0.5 & 0.25 & 0.0625 & 0.0625 & 0.0625 & 0.0625 \\
0.25 & 0.5 & 0.0625 & 0.0625 & 0.0625 & 0.0625 \\
0.0625 & 0.0625 & 0.5 & 0.25 & 0.0625 & 0.0625 \\
0.0625 & 0.0625 & 0.25 & 0.5 & 0.0625 & 0.0625 \\
0.0625 & 0.0625 & 0.0625 & 0.0625 & 0.5 & 0.25 \\
0.0625 & 0.0625 & 0.0625 & 0.0625 & 0.25 & 0.5
\end{pmatrix}.
$$

\noindent We considered two covariates:
\begin{enumerate} \item[$\bullet$]
$X_{ij1}$ a binary variable such as $P(X_{ij1}=0)=P(X_{ij1}=1)=1/2$
\item[$\bullet$] $X_{ij2}$ a uniform variable over $[0,4]$.
\end{enumerate}

\noindent Given these covariates, the hazard function of $T_{ij}$ is
$$\lambda(t|X_{ij1},X_{ij2})=\lambda_{0}(t) e^{\beta_1X_{ij1} + r(t-X_{ij2})}.$$

\noindent Here, we set $\lambda_0(t)=\alpha\lambda t^{\alpha-1}$, $\lambda=1/1000$, $\alpha=4$, $\beta_1=1$ and \[
r(u)= 0.55
+ 0.35 \sin\!\left(\frac{2\pi u}{1.8}\right) e^{-0.35 u}
+ 0.15 \frac{u}{4}.
\]
\noindent The proband's ages at entry $\{a_i, i=1,\cdots,I\}$ were generated from a Weibull distribution such that $E[a_i]=4$ and $Var[a_i]=2$, whereas the censoring times $\{c_{ij},i=1,\cdots,I, j=1\cdots,6\}$ were generated from a uniform distribution over $[0,10]$. Furthermore, we set the heritability parameter to $h=1/2$.

\noindent Given the covariates $(X_{ij1},X_{ij2})$, the proband ages at entry $a_i$ and the censoring times $c_{ij}$, we generated the survival data $(y_{ij},\delta_{ij})$ with a selection bias following these steps:\\

\noindent For $i=1,\cdots,I$
\begin{enumerate}
\item[$\bullet$] Generate $v_{i}$ from a uniform distribution over $[0,1]$

\item[$\bullet$] Solve $$P(T_{i0}>t \mid T_{i0}<a_i, X_{i01}, X_{i02}) =  \frac{S(t|X_{i01}, X_{i02})-S(a|X_{i01}, X_{i02})}{1-S(a|X_{i01}, X_{i02})}= v_i$$ in $t$ to obtain $t_{i0}$.

\item[$\bullet$] Compute $u_{i0}=S(t_{i0}|X_{i01},X_{i02})$ and $z_{i0}=\Phi^{-1}(u_{i0})$.
\item[$\bullet$] Generate $\{z_{i1},\cdots,z_{in_i}\}$ from a conditional multivariate normal distribution with mean 0 and covariance matrix equal to $V^{(i)}$ given $Z_{i0}=z_{i0}$.
\item[$\bullet$] For $j=1,\cdots,n_i$
\begin{enumerate}
\item[{\it (i)}] Compute $u_{ij} = \Phi(z_{ij})$.

\item[{\it (ii)}] Solve $S(t|X_{ij1},X_{ij2})=u_{ij}$ in $t$ to obtain $t_{ij}$
\item[{\it (iii)}] Set $y_{ij}=\mbox{min}(t_{ij},c_{ij})$ and $\delta_{ij}=1_{\{t_{ij}<c_{ij}\}}$.
\end{enumerate}
\end{enumerate}

\noindent We generated 200 datasets following the algorithm described above with $I=200$ families. For each dataset, we estimated the parameters $\Theta$, along with their variances. The results obtained with $\gamma \in \{0, 0.05, 0.2, 0.8\}$ are reported in Tables 2 and 3. In Figure 6, we report the mean Splines-based estimate of the function $r$, along with the true one. The four plots correspond to $\gamma \in \{0, 0.05, 0.2, 0.8\}$.


\begin{table}[H]
\centering
\small
\renewcommand{\arraystretch}{1}
\setlength{\tabcolsep}{2.5pt}
\resizebox{1.1\textwidth}{!}{%
\begin{tabular}{|c|c|cccc|cccc|}
\hline
 &  & \multicolumn{4}{c|}{$\gamma=0$}
 & \multicolumn{4}{c|}{$\gamma=0.05$} \\
\hline
Parameters & Real 
& Bias & Empirical & Asymptotic & Cover 
& Bias & Empirical & Asymptotic & Cover \\
 & Value &  & var. & var. & rate
 &  & var. & var. & rate \\
\hline

$\log(\alpha)$ & 1.3863  &0.0057	&0.0017	& 0.0015 & 92.50\% 	& 0.0065 &	0.0017 & 0.0015	& 94.50\% \\
$\log(\lambda)$ & -6.9078 & -0.0330	&0.0563	&0.0527	&94.00\%	&-0.0356	&0.0560	&0.0519	& 94.50\% \\
$\beta_{\text{gene}}$ & 1.0000 & 0.0025	&0.0025&	0.0028	&95.50\%	&0.0072	&0.0025	&0.0027	&96.00\% \\
$h$ & 0.5000 & -0.0028&	0.0128&	0.0098	&90.50\%&-0.0083	&0.0128	&0.0096	&90.50\% \\

\hline
\end{tabular}%
}

\caption{Simulation results for the B-spline model ($n_{\text{basis}}=10$, order $=4$): empirical means, empirical variances, asymptotic variances, and coverage rates under different values of the penalty parameter $\gamma$. The parameters $b_1, b_2, \ldots, b_{10}$ denote the coefficients of the B-spline basis functions.}
\label{tab:simulation_bspline}
\end{table}


\begin{table}[H]
\centering
\small
\renewcommand{\arraystretch}{1}
\setlength{\tabcolsep}{2.5pt}
\resizebox{1.1\textwidth}{!}{%
\begin{tabular}{|c|c|cccc|cccc|}
\hline
 &  & \multicolumn{4}{c|}{$\gamma=0.2$}
 & \multicolumn{4}{c|}{$\gamma=0.08$} \\
\hline
Parameters & Real 
& Bias & Empirical & Asymptotic & Cover 
& Bias & Empirical & Asymptotic & Cover \\
 & Value &  & var. & var. & rate
 &  & var. & var. & rate \\
\hline
$\log(\alpha)$ & 1.3863  &0.0068	&0.0017	&0.0015	&93.00\%	& 0.0063&	0.0017	&0.0015&	94.00\%\\
$\log(\lambda)$ & -6.9078 &-0.0407	&0.0563	&0.0527	&93.50\%	&	-0.0380&	0.0560	&0.0528	&93.50\% \\
$\beta_{\text{gene}}$ & 1.0000 & 0.0077	&0.0026	&0.0028	&96.00\%	&	0.0076	&0.0026	&0.0027&	95.00\%\\
$h$ & 0.5000 & -0.0099	&0.0126	&0.0097	&91.00\%	&	-0.0100	&0.0126	&0.0097&	91.00\% \\
\hline
\end{tabular}%
}

\caption{Simulation results for the B-spline model ($n_{\text{basis}}=10$, order $=4$): empirical means, empirical variances, asymptotic variances, and coverage rates under different values of the penalty parameter $\gamma$. The parameters $b_1, b_2, \ldots, b_{10}$ denote the coefficients of the B-spline basis functions.}
\label{tab:simulation_bspline_020_008}
\end{table}

\begin{figure}[H]
   \centering
 \includegraphics[scale=0.79]{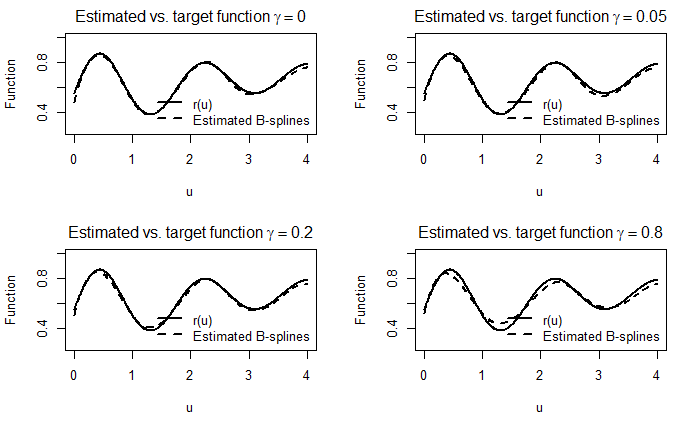}
 \caption{ Estimated vs. True function $r$}
\end{figure}

\noindent From these Tables and these figures, we notice that the proposed iterative algorithm estimated the parameters of the model \textcolor{black}{in the presence of ascertainment through the proband} without bias. Moreover, the sandwich variance estimator yields reliable results.

\noindent When the data is correctly specified, good precision can be achieved without penalization, whereas larger values of $\gamma$ may lead to biased estimates.

\noindent The heritability parameter $h$ is estimated with relatively good precision. This is explained by the nature of \( h \), which is a variance component. Therefore, $\hat h$ would follow a chi-squared type of distribution when $I$ is small. The sampling distribution of $h$ is well approximated by a normal distributrion only with large sample sizes.

\section{Conclusion}

\noindent The main methodological contribution of this work is the developpement of an iterative algortihm to estimate the marginal and association parameters in a copula model for familial survival data in the presence of a time-varying covariate and a selection bias.

\noindent \textcolor{black}{From our simulation study, we concluded that the proposed estimation procedure yielded unbiased estimates of the model parameters and of their variances.}

\noindent This study is motived by the analysis of {\it BRCA1} families data from BCFR. The age at RRSO induces a time-varying covariate. Choi et al. (2021) considered a parametric form for the effect of this covariate on BC onset and noticed that BC risk estimates could be quite sensitive to the specification of this parametric form. In this work, we considered a flexible approach based on B-Splines and obtained relatively high variable estimates. This is due to the fact that we only have 26 observed events in the data at hand to estimate the coefficients of the B-splines. This motivated us to consider a penalized log-likelihood approach with a smoothing parameter $\gamma$ to reduce the variance of the obtained estimates. In our application, we applied a rule of thumb to select the smoothing parameter $\gamma$. Investigating numerical criteria to select the optimal $\gamma$ and considering alternative modelling approaches of the effect of RRSO on the age of BC onset is subject of ongoing research.

\noindent The developped model enabled us to define and estimate personalised risk functions that incorporate exact familial cancer history \textcolor{black}{as well as the family history of prophylactic surgery, i.e., RRSO.} Such estimates are of prime importance for practitioners to efficiently manage enhanced cancer screening programs \citep{jacobs2021predicting}. \textcolor{black}{To our knowledge, such estimates are not yet implemented in BC risk prediction models such as BOADICEA \citep{lee2019} or BCRAT \citep{gail1989}.} Similarly, these models can be further extended to identify people with deleterious pathogenic variants and to estimate the optimal age at RRSO or other interventions. These topics are subject of ongoing research.

\noindent The illustration of the developed iterative algorithm was limited to the case of a Weibull baseline hazard and a Gaussian copula. Other parametric distributions and semi-parametric approaches such as the piecewise constant hazard can be considered to model the marginal distribution of the survival times without loss of generality. Moreover, any elliptic copula parametrized by a correlation copula is suitable for this context. From our long experience with multivariate survival data, the results are typically robust to the specification of the parametric form of the copula. However, the obtained outcomes may vary with the specification of the marginal distribution. One may rely on criteria such as AIC or BIC based on $l^{(1)}(\theta|h,O)$ to select the most suitable marginal model.

\noindent  The iterative algorithm presented in this work was implemented in \texttt{R}. The code to reproduce the simulations is hosted in the repository \url{https://github.com/siseb-create/copula-familial-survival-simulation}. The computation time of the algorithm has been significantly reduced by considering an efficient calculation of the integrals present in the log-likelihood functions. This increases the practical usefulness of the proposed method. 

\noindent This work was limited to the case of survival data. Choi et al. (2021) considered a competing risks setting using frailties. Su et al (2020) considered a copula model for clustered competing risks. Adapting these latter approached and the inference procedure to the presence of a selection bias is subject of ongoing research.

\section{Acknowledgements}
This work was supported by grant U01 CA164920 from the USA National Cancer Institute. The content of this manuscript does not necessarily reflect the views or policies of the National Cancer Institute or any of the collaborating centers in the Breast Cancer Family Registry (BCFR), nor does mention of trade names, commercial products, or organizations imply endorsement by the USA Government or the BCFR.

\section*{Appendix}
\subsection*{Appendix A: Elements of the Bivariable log-likelihood function}
We distinguish four cases of pair contributions to the likelihood. For any family $i$ and $j\neq k$ we have:\\
\begin{enumerate}
   \item  If $\delta_{ij}=0$ and $\delta_{ik}=0$, then the contribution of the pair $(j, k)$ is:
\begin{align}
  &  \dfrac{\partial}{\partial y_{i0}} P(T_{ij}>y_{ij}, T_{ik}>y_{ik}|T_{i0}>y_{i0},X_{ij}(y_{ij}), X_{ik}(y_{ik}), X_{i0}(y_{i0})) \nonumber\\
    &\propto D_{\textcolor{black}{100}}\{S(y_{ij}|X_{ij}(y_{ij})), S(y_{ik}|X_{ik}(y_{ik})), S(y_{i0}|X_{i0}(y_{i0}))\}.
\end{align}
   \item  If $\delta_{ij}=0$ and $\delta_{ik}=1$, then the contribution of the pair $(j, k)$ is
\begin{align}
  &\dfrac{\partial^2}{\partial y_{ik} \partial y_{i0}} P(T_{ij}>y_{ij},T_{ik}>y_{ik}|T_{i0}>y_{i0}, X_{ij}(y_{ij}), X_{ik}(y_{ik}),X_{i0}(y_{i0})) \nonumber\\ 
  &\propto D_{\textcolor{black}{101}}\{S(y_{ij}|X_{ij}(y_{ij})), S(y_{ik}|X_{ik}(y_{ik})), S(y_{i0}|X_{i0}(y_{i0})) \}. 
\end{align}

 \item If $\delta_{ij}=1$ and $\delta_{ik}=0$, then the contribution of the pair $(j, k)$ is 
\begin{align}
  &\dfrac{\partial^2}{\partial y_{ij} \partial y_{i0}} P(T_{ij}>y_{ij},T_{ik}>y_{ik}|T_{i0}>y_{i0}, X_{ij}(y_{ij}), X_{ik}(y_{ik}),X_{i0}(y_{i0})) \nonumber\\ 
  &\propto D_{\textcolor{black}{110}}\{S(y_{ij}|X_{ij}(y_{ij})), S(y_{ik}|X_{ik}(y_{ik})), S(y_{i0}|X_{i0}(y_{i0}))\}. 
\end{align}

   \item If $\delta_{ij}=1$ and $\delta_{il}=1$, then the contribution of the pair $(j, l)$ is

   \begin{align}
  &\dfrac{\partial^3}{\partial y_{ij} \partial y_{ik} \partial y_{i0}} P(T_{ij}>y_{ij},T_{ik}>y_{ik}|T_{i0}>y_{i0}, X_{ij}(y_{ij}), X_{ik}(y_{ik}),X_{i0}(y_{i0})) \nonumber\\ 
  &\propto D_{111}\{S(y_{ij}|X_{ij}(y_{ij})), S(y_{ik}|X_{ik}(y_{ik})), S(y_{i0}|X_{i0}(y_{i0}))\}.  
\end{align}

\end{enumerate}
The log-\textcolor{black}{likelihood} contributions from each of the pairs of nonproband individuals in family $i$ can be written as:

\begin{align*}
    l^{(2)}_{\textcolor{black}{\{i,jk|j \neq k\}}}(h|\theta)&=\log  \dfrac{\partial}{\partial y_{i0}} P(T_{ij}>y_{ij}, T_{ik}>y_{ik}|T_{i0}>y_{i0},X_{ij}(y_{ij}), X_{ik}(y_{ik}), X_{i0}(y_{i0}))\mathbf{1}_{\{\delta_{ij}=0, \ \delta_{ik}=0\}} + \\
    &+\log \dfrac{\partial^2}{\partial y_{ik} \partial y_{i0}} P(T_{ij}>y_{ij},T_{ik}>y_{ik}|T_{i0}>y_{i0}, X_{ij}(y_{ij}), X_{ik}(y_{ik}),X_{i0}(y_{i0})) \mathbf{1}_{\{\delta_{ij}=0,\  \delta_{ik}=1\}}  +\\ 
    &+\log \dfrac{\partial^2}{\partial y_{ij} \partial y_{i0}} P(T_{ij}>y_{ij},T_{ik}>y_{ik}|T_{i0}>y_{i0}, X_{ij}(y_{ij}), X_{ik}(y_{ik}),X_{i0}(y_{i0})) \mathbf{1}_{\{\delta_{ij}=1,\  \delta_{ik}=0\}} + \\
    &+\log \dfrac{\partial^3}{\partial y_{ij} \partial y_{ik} \partial y_{i0}} P(T_{ij}>y_{ij},T_{ik}>y_{ik}|T_{i0}>y_{i0}, X_{ij}(y_{ij}), X_{ik}(y_{ik}),X_{i0}(y_{i0})) \mathbf{1}_{\{\delta_{ij}=1,\  \delta_{ik}=1\}}\\
    &=\log\left[ D_{100}\left\{S(t_{i0}|X_{i0}),S(t_{ij}|X_{ij}),S(t_{ik}|X_{ik})\right\}\right]\mathbf{1}_{\{\delta_{ij}=0, \ \delta_{ik}=0\}}\\
    &+\log\left[ D_{101}\left\{S(t_{i0}|X_{i0}),S(t_{ij}|X_{ij}),S(t_{ik}|X_{ik})\right\}\right]\mathbf{1}_{\{\delta_{ij}=0, \ \delta_{ik}=1\}}\\
    &+\log\left[ D_{110}\left\{S(t_{i0}|X_{i0}),S(t_{ij}|X_{ij}),S(t_{ik}|X_{ik})\right\}\right]\mathbf{1}_{\{\delta_{ij}=1, \ \delta_{ik}=0\}}\\
    &+\log\left[ D_{111}\left\{S(t_{i0}|X_{i0}),S(t_{ij}|X_{ij}),S(t_{ik}|X_{ik})\right\}\right]\mathbf{1}_{\{\delta_{ij}=1, \ \delta_{ik}=1\}}
\end{align*}
Then, by posing
\begin{align*} 
&\log\left[ D_{1lm}\left\{S(t_{i0}|X_{i0}),S(t_{ij}|X_{ij}),S(t_{ik}|X_{ik})\right\}\right]\mathbf{1}_{\{\delta_{ij}=l, \ \delta_{ik}=m\}}\\ 
&=\log\left[ D_{1\delta_j\delta_k}\left\{S(t_{i0}|X_{i0}),S(t_{ij}|X_{ij}),S(t_{ik}|X_{ik})\right\}\right]
\end{align*}
with $l \in\{0,1\}$ and $m\in\{0,1\}$, the likelihood of a family $i$ can to be written in the form of $$ \sum\limits_{1\le j < k  \le n_i} \log\left[ D_{1\delta_j\delta_k}\left\{S(t_{i0}|X_{i0}),S(t_{ij}|X_{ij}),S(t_{ik}|X_{ik})\right\}\right].$$ So, the log-likelihood of I families is:
$$ \sum\limits_{i=1}^I \sum\limits_{1\le j < k  \le n_i} \log\left[ D_{1\delta_j\delta_k}\left\{S(t_{i0}|X_{i0}),S(t_{ij}|X_{ij}),S(t_{ik}|X_{ik})\right\}\right].$$

\newpage
\subsection*{ Appendix B : Database structure}

\begin{table}[H]
\centering
\caption{Distribution of family sizes among study families with probands diagnosed with breast cancer prior to study entry}
\begin{tabular}{ccc}
\toprule
Family size & Number of families & Total individuals \\
\midrule
1  & 35 & 35 \\
2  & 46 & 92 \\
3  & 54 & 162 \\
4  & 44 & 176 \\
5  & 47 & 235 \\
6  & 36 & 216 \\
7  & 30 & 210 \\
8  & 17 & 136 \\
9  & 7  & 63 \\
10 & 8  & 80 \\
11 & 10 & 110 \\
12 & 2  & 24 \\
13 & 4  & 52 \\
14 & 3  & 42 \\
15 & 2  & 30 \\
43 & 1  & 43 \\
\midrule
Total &  & 1706 \\
\bottomrule
\end{tabular}
\end{table}

\newpage
\subsection*{Appendix D}

\begin{figure}[H]
\centering
 \includegraphics[scale=0.8]{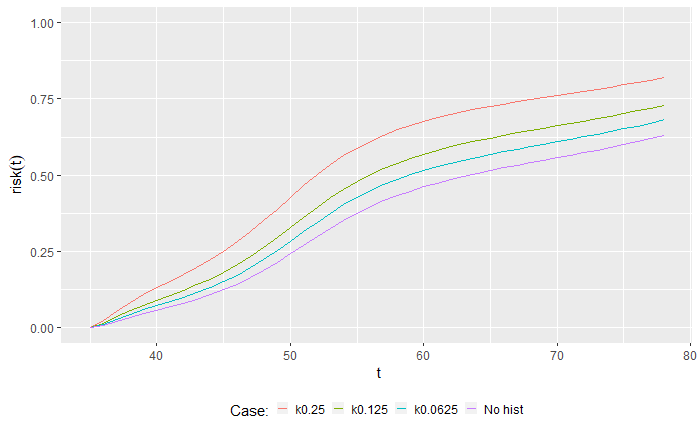}
 \caption{Estimate of $\text{risk}(t)$  
 as a function of $t$ when $X_{ij1} = 32$, $X_{ik1} = \infty$, and when $X_{ij2} =1,\  X_{ik2} = 0$.}  
\end{figure}

\begin{figure}[H]
\centering
 \includegraphics[scale=0.8]{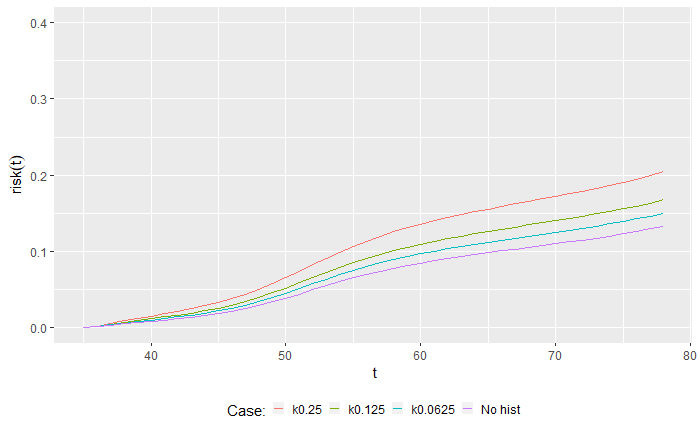}
  \caption{Estimate of $\text{risk}(t)$  
  as a function of $t$ when $X_{i1} = 32$, $X_{ik1} = \infty$, and when $X_{ij2} =0,\  X_{ik2} = 1$.}
\end{figure}

\begin{figure}[H]
   \centering
 \includegraphics[scale=0.8]{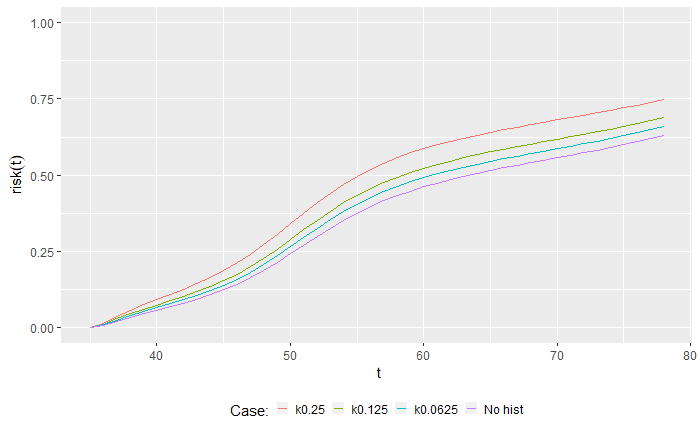}
 \caption{Estimate of $\text{risk}(t)$  
 as a function of $t$ when $X_{ij1} = 32$, $X_{ik1} = \infty$, and when $X_{ij2} = X_{ik2} = 1$.}
\end{figure}

\begin{figure}[H]
   \centering
 \includegraphics[scale=0.8]{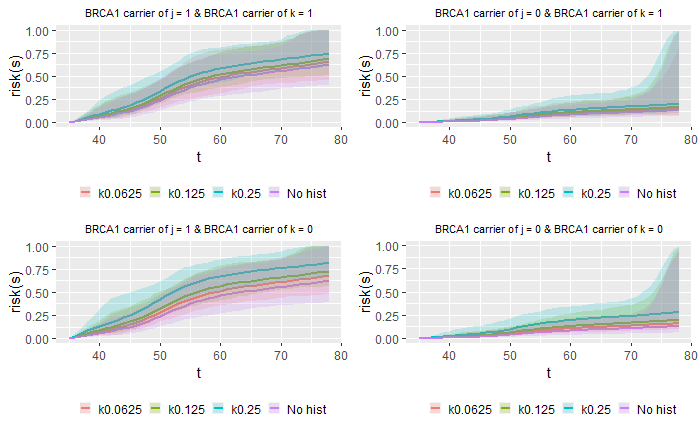}
 \caption{Estimate of confidence interval of $\text{risk}(t)$ as a function of $t$ when $X_{ij1} = 32$, $X_{ik1} = \infty$.}
\end{figure}


\subsection*{Appendix E}

\begin{figure}[H]
   \centering
 \includegraphics[scale=0.8]{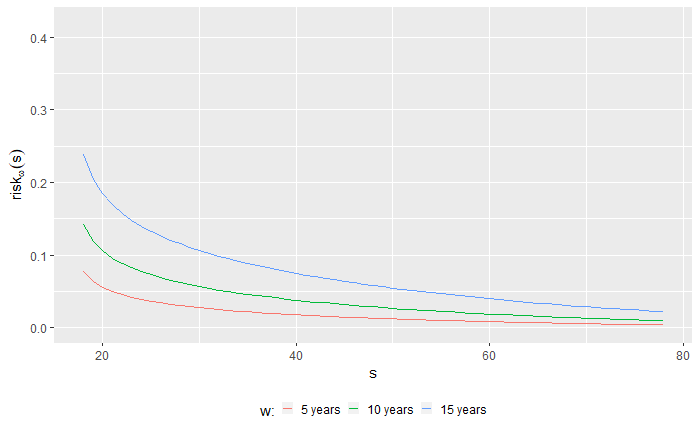}
 \caption{Estimate of $\text{risk}_\omega(s)$  
 as a function of $s$ when $X_{ij1} = 32$, $X_{ik1} = \infty$, and $X_{ij2} =0,\  X_{ik2} = 1$. The three curves correspond to $\omega = 5$, $10$, and $15$ years, respectively.}
\end{figure}

\begin{figure}[H]
   \centering
 \includegraphics[scale=0.78]{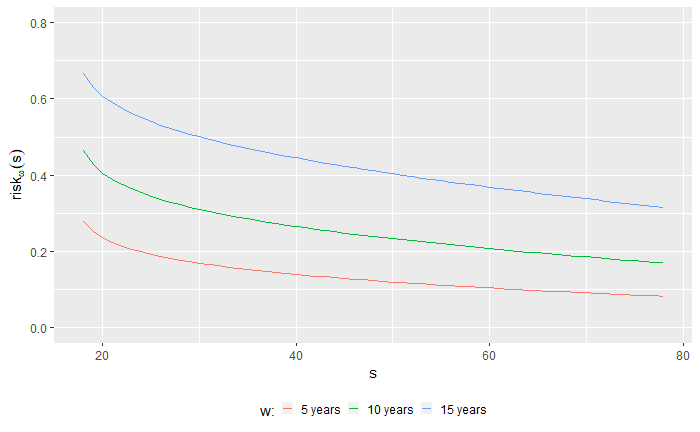}
 \caption{Estimate of $\text{risk}_\omega(s)$  
 as a function of $s$ when $X_{ij1} = 32$, $X_{ik1} = \infty$, and $X_{ij2} =1,\  X_{ik2} = 0$. The three curves correspond to $\omega = 5$, $10$, and $15$ years, respectively.}
\end{figure}

\begin{figure}[H]
   \centering
 \includegraphics[scale=0.8]{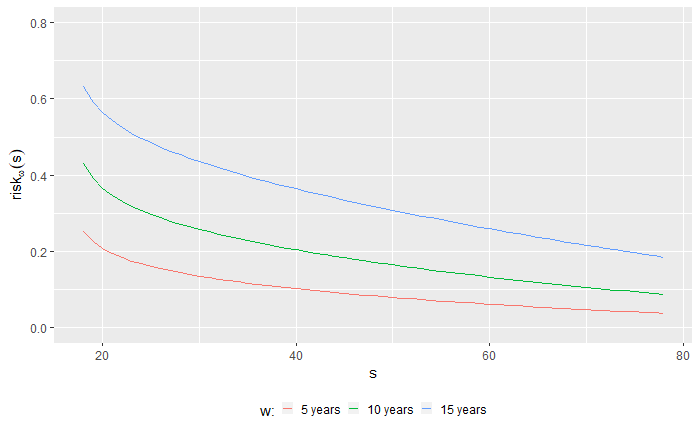}
 \caption{Estimate of $\text{risk}_\omega(s)$  
 as a function of $s$ when $X_{ij1} = 32$, $X_{ik1} = \infty$, and $X_{ij2} = X_{ik2} = 1$. The three curves correspond to $\omega = 5$, $10$, and $15$ years, respectively.}
\end{figure}

\begin{figure}[H]
   \centering
 \includegraphics[scale=0.8]{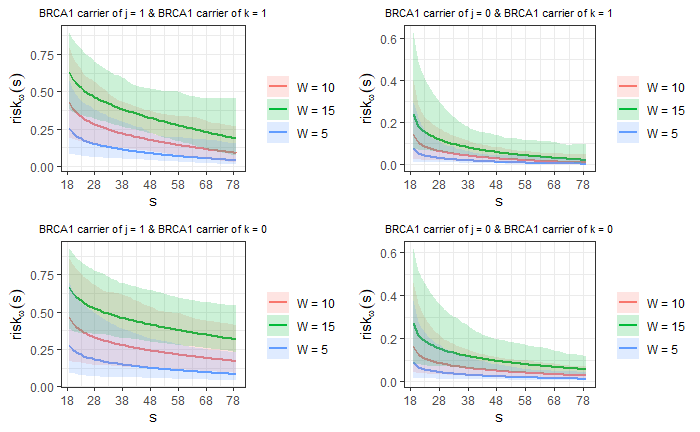}
 \caption{Estimate of confidence interval of $\text{risk}_\omega(s)$ as a function of $s$ when $X_{ij1} = 32$, $X_{ik1} = \infty$. The three curves correspond to $\omega = 5$, $10$, and $15$ years, respectively.}
\end{figure}

\subsection*{Appendix F}

\begin{figure}[H]
  \centering
 \includegraphics[scale=0.75]{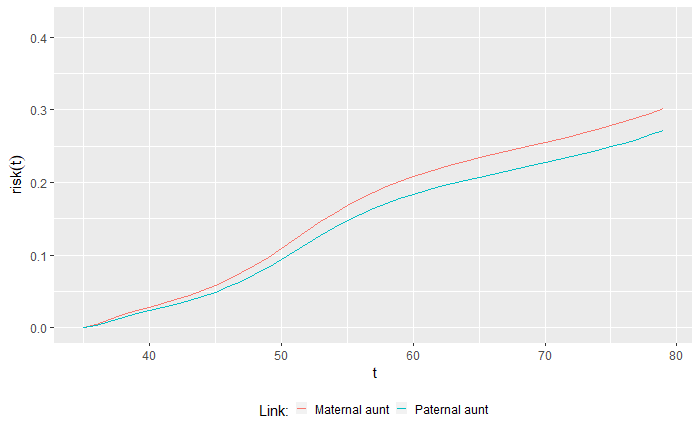}
  \caption{Estimate of $\text{risk}(t)$ 
  as a function of $t$ when $X_{ij1} = 32$, $X_{ik1} = X_{il1} = \infty$, and $X_{ij2} =0,\  X_{ik2} =0,\  X_{il2} = 1$. The orange curves refer to estimates when the aunt is maternal, whereas the blue ones correspond to the case of a paternal aunt.}
\end{figure}

\begin{figure}[H]
   \centering
 \includegraphics[scale=0.75]{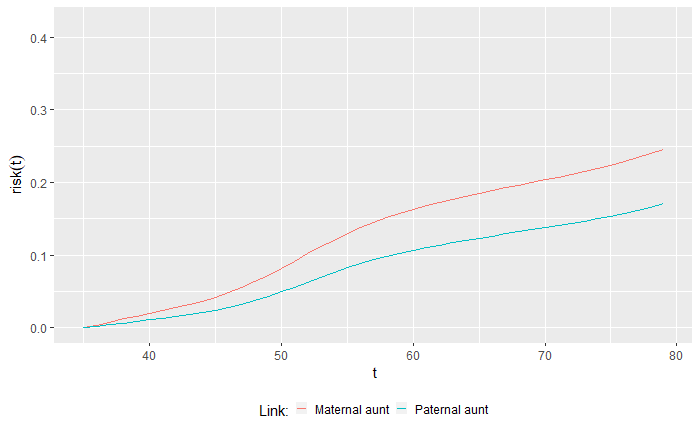}
 \caption{Estimate of $\text{risk}(t)$
 as a function of $t$ when $X_{ij1} = 32$, $X_{ik1} = X_{il1} = \infty$, and $X_{ij2} =0,\  X_{ik2} =1,\  X_{il2} = 0$. The orange curves refer to estimates when the aunt is maternal, whereas the blue ones correspond to the case of a paternal aunt.}
\end{figure}

\begin{figure}[H]
   \centering
 \includegraphics[scale=0.8]{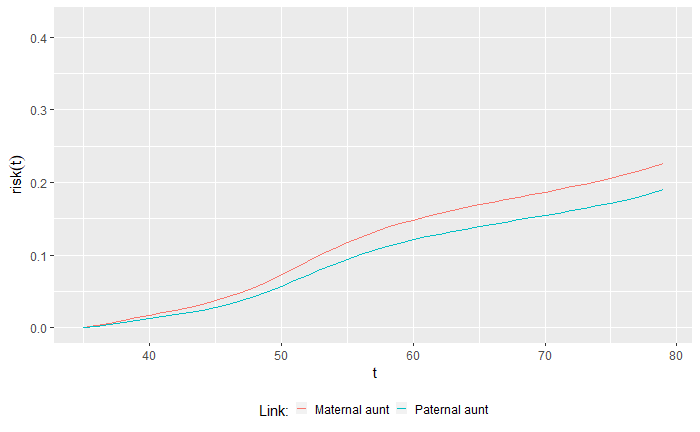}
 \caption{Estimate of $\text{risk}(t)$  
 as a function of $t$ when $X_{ij1} = 32$, $X_{ik1} = X_{il1} = \infty$, and $X_{ij2} =0,\  X_{ik2} =1,\  X_{il2} = 1$. The orange curves refer to estimates when the aunt is maternal, whereas the blue ones correspond to the case of a paternal aunt.}
\end{figure}

\begin{figure}[H]
   \centering
 \includegraphics[scale=0.8]{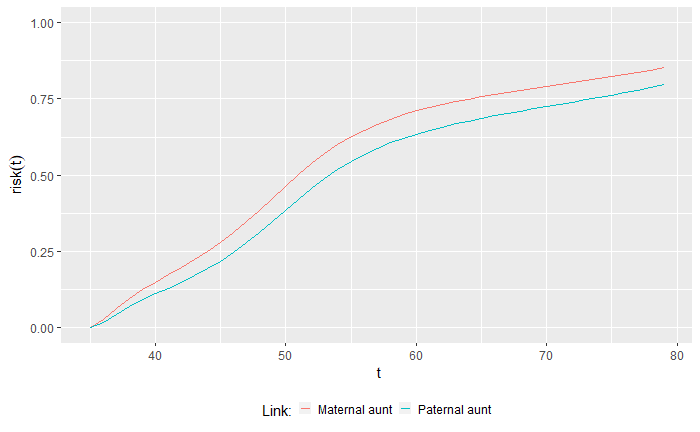}
 \caption{Estimate of $\text{risk}(t)$ 
 as a function of $t$ when $X_{ij1} = 32$, $X_{ik1} = X_{il1} = \infty$, and $X_{ij2} =1,\  X_{ik2} =0,\  X_{il2} = 0$. The orange curves refer to estimates when the aunt is maternal, whereas the blue ones correspond to the case of a paternal aunt.}
\end{figure}

\begin{figure}[H]
  \centering
 \includegraphics[scale=0.8]{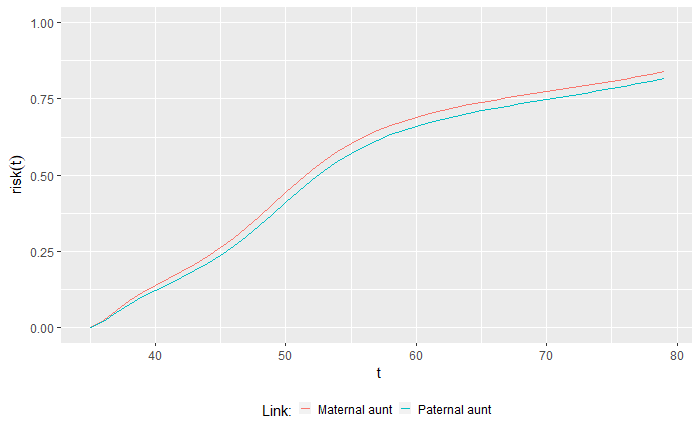}
  \caption{Estimate of $\text{risk}(t)$  
  as a function of $t$ when $X_{ij1} = 32$, $X_{ik1} = X_{il1} = \infty$, and $X_{ij2} =1,\  X_{ik2} =0,\  X_{il2} = 1$. The orange curves refer to estimates when the aunt is maternal, whereas the blue ones correspond to the case of a paternal aunt.}
\end{figure}

\begin{figure}[H]
   \centering
 \includegraphics[scale=0.8]{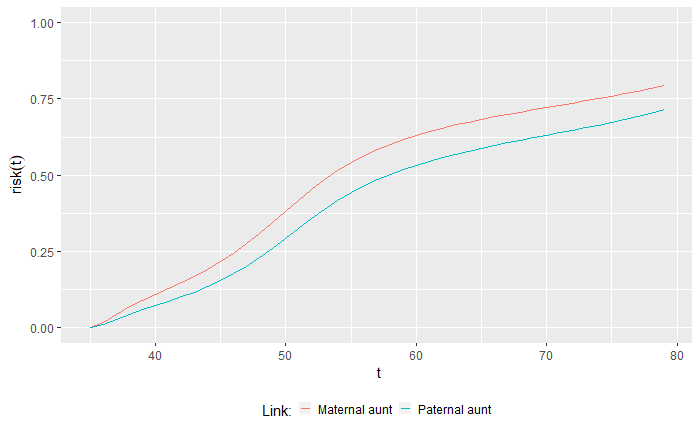}
 \caption{Estimate of $\text{risk}(t)$  
 as a function of $t$ when $X_{ij1} = 32$, $X_{ik1} = X_{il1} = \infty$, and $X_{ij2} =1,\  X_{ik2} =1,\  X_{il2} = 0$. The orange curves refer to estimates when the aunt is maternal, whereas the blue ones correspond to the case of a paternal aunt.}
\end{figure}

\begin{figure}[H]
   \centering
 \includegraphics[scale=0.8]{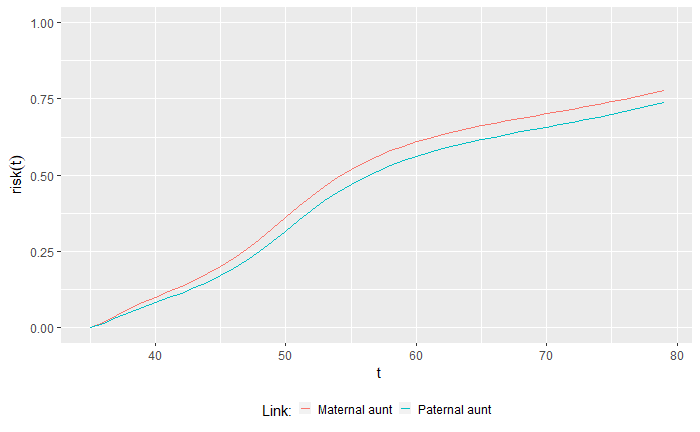}
 \caption{Estimate of $\text{risk}(t)$  
 as a function of $t$ when $X_{ij1} = 32$, $X_{ik1} = X_{il1} = \infty$, and $X_{ij2} =1,\  X_{ik2} =1,\  X_{il2} = 1$. The orange curves refer to estimates when the aunt is maternal, whereas the blue ones correspond to the case of a paternal aunt.}
\end{figure}

\begin{figure}[H]
   \centering
 \includegraphics[scale=0.8]{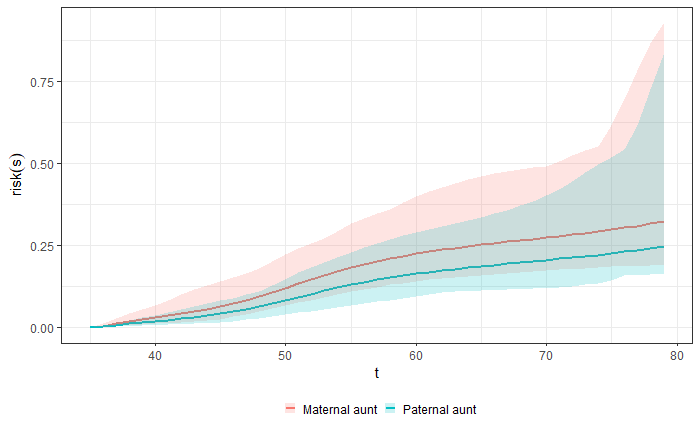}
 \caption{Estimate of confidence interval of $\text{risk}(t)$  
 as a function of $t$ when $X_{ij1} = 32$, $X_{ik1} = X_{il1} = \infty$, and $X_{ij2} =0,\  X_{ik2} =0,\  X_{il2} = 0$. The orange curves refer to estimates when the aunt is maternal, whereas the blue ones correspond to the case of a paternal aunt.}
\end{figure}

\subsection*{Appendix H:Appendix H. Pedigree structures: graphical representations}


\begin{figure}[h]
\centering
\begin{tikzpicture}[
  x=1.1cm, y=1.1cm,
  female/.style={circle, draw, line width=0.6pt, minimum size=5.8mm, inner sep=0pt},
  rel/.style={line width=0.6pt},
  lab/.style={font=\small}
]

\node[female] (j) at (0,0) {};
\node[female] (k) at (3,0) {};

\draw[rel] (j) -- node[lab, above] {$\phi^{(i)}_{jk}$} (k);

\node[lab, below=2pt of j] {$j$};
\node[lab, below=2pt of k] {$k$};

\end{tikzpicture}
\caption{Schematic representation of a pair $(j,k)$ from family $i$ with kinship coefficient $\phi^{(i)}_{jk}$, where the cancer history of $k$ is used to update the risk for $j$.}
\label{fig:pair_generic_phi}
\end{figure}
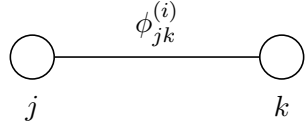


\begin{figure}[H]
\centering
\begin{tikzpicture}[
  x=1.1cm, y=1.1cm,
  female/.style={circle, draw, line width=0.6pt, minimum size=5.8mm, inner sep=0pt},
  male/.style={rectangle, draw, line width=0.6pt, minimum size=5.8mm, inner sep=0pt},
  rel/.style={line width=0.6pt},
  lab/.style={font=\small}
]

\node[male]   (F) at (0,2.2) {};
\node[female] (M) at (1,2.2) {};
\draw[rel] (F) -- (M);

\coordinate (sibmid) at (0.5,1.75);
\draw[rel] (0.5,2.2) -- (sibmid);
\draw[rel] (-0.8,1.75) -- (1.8,1.75);

\node[female] (j) at (-0.8,1.2) {}; \node[lab, below=2pt of j] {$j$};
\node[female] (k) at ( 1.8,1.2) {}; \node[lab, below=2pt of k] {$k$};

\draw[rel] (j) -- (-0.8,1.75);
\draw[rel] (k) -- ( 1.8,1.75);

\end{tikzpicture}
\caption{Pedigree for a pair $(j,k)$ in family $i$ with $\phi^{(i)}_{jk}=0.25$ (full sisters). Individual $j$ is followed up to age $t_{ij}=35$, remains cancer-free, and undergoes oophorectomy at $X_{ij1}=32$. Individual $k$ is diagnosed at age $T_{ik}=s$ and never undergoes oophorectomy ($X_{ik1}=\infty$).}
\label{fig:ped_phi05}
\end{figure}
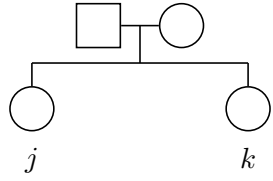

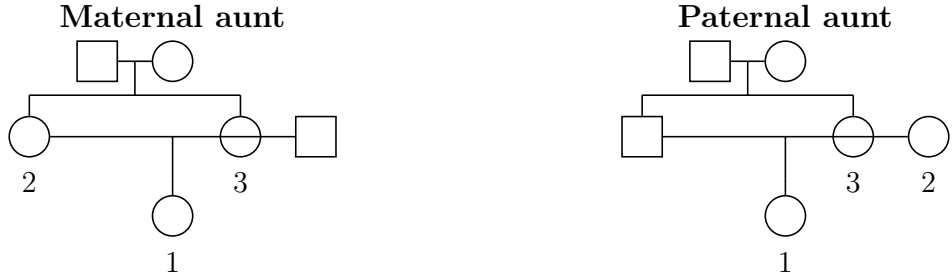
\begin{figure}[H]
\centering

\begin{minipage}[t]{0.48\textwidth}
\centering
\textbf{Maternal aunt}\\[4pt]
\begin{tikzpicture}[
  scale=0.95, transform shape,
  x=1.05cm, y=1.05cm,
  female/.style={circle, draw, line width=0.6pt, minimum size=5.6mm, inner sep=0pt},
  male/.style={rectangle, draw, line width=0.6pt, minimum size=5.6mm, inner sep=0pt},
  rel/.style={line width=0.6pt}
]

\node[male]   (gf) at (0,3) {};
\node[female] (gm) at (1,3) {};
\draw[rel] (gf)--(gm);

\coordinate (sibmid) at (0.5,2.55);
\draw[rel] (0.5,3)--(sibmid);
\draw[rel] (-0.9,2.55)--(1.9,2.55);

\node[female] (m2) at (-0.9,2) {};
\node[female] (a3) at ( 1.9,2) {};
\draw[rel] (m2)--(-0.9,2.55);
\draw[rel] (a3)--( 1.9,2.55);

\node[male] (fa) at (2.9,2) {};
\draw[rel] (m2)--(fa);

\coordinate (cp) at ($(m2)!0.5!(fa)$);
\draw[rel] (cp)--($(cp)+(0,-0.55)$);
\node[female] (c1) at ($(cp)+(0,-1.05)$) {};
\draw[rel] (c1)--($(cp)+(0,-0.55)$);

\node[below=2pt of m2] {$2$};
\node[below=2pt of a3] {$3$};
\node[below=2pt of c1] {$1$};

\end{tikzpicture}
\end{minipage}
\hfill
\begin{minipage}[t]{0.48\textwidth}
\centering
\textbf{Paternal aunt}\\[4pt]
\begin{tikzpicture}[
  scale=0.95, transform shape,
  x=1.05cm, y=1.05cm,
  female/.style={circle, draw, line width=0.6pt, minimum size=5.6mm, inner sep=0pt},
  male/.style={rectangle, draw, line width=0.6pt, minimum size=5.6mm, inner sep=0pt},
  rel/.style={line width=0.6pt}
]

\node[male]   (gf) at (0,3) {};
\node[female] (gm) at (1,3) {};
\draw[rel] (gf)--(gm);

\coordinate (sibmid) at (0.5,2.55);
\draw[rel] (0.5,3)--(sibmid);
\draw[rel] (-0.9,2.55)--(1.9,2.55);

\node[male]   (fa) at (-0.9,2) {};
\node[female] (a3) at ( 1.9,2) {};
\draw[rel] (fa)--(-0.9,2.55);
\draw[rel] (a3)--( 1.9,2.55);

\node[female] (m2) at (2.9,2) {};
\draw[rel] (fa)--(m2);

\coordinate (cp) at ($(fa)!0.5!(m2)$);
\draw[rel] (cp)--($(cp)+(0,-0.55)$);
\node[female] (c1) at ($(cp)+(0,-1.05)$) {};
\draw[rel] (c1)--($(cp)+(0,-0.55)$);

\node[below=2pt of a3] {$3$};
\node[below=2pt of m2] {$2$};
\node[below=2pt of c1] {$1$};

\end{tikzpicture}
\end{minipage}

\caption{Pedigree structures inducing the two relationship matrices among the observed individuals $1$ (child), $2$ (mother), and $3$ (aunt). 
}
\label{fig:aunt_cases}
\end{figure}


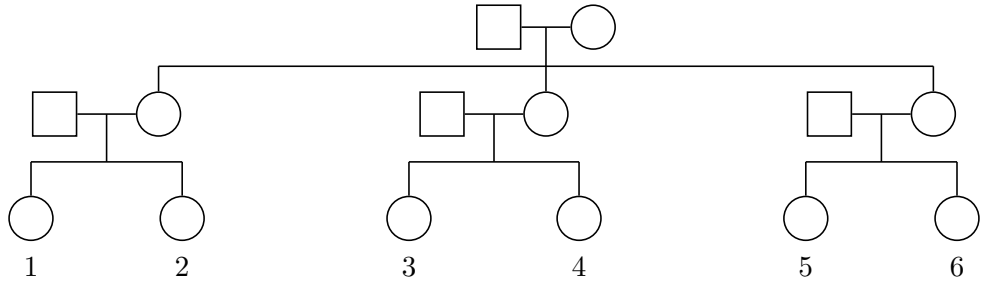
\begin{figure}[H]
\centering
\begin{tikzpicture}[
  x=1.25cm, y=1.15cm,
  female/.style={circle, draw, line width=0.6pt, minimum size=5.8mm, inner sep=0pt},
  male/.style={rectangle, draw, line width=0.6pt, minimum size=5.8mm, inner sep=0pt},
  rel/.style={line width=0.6pt},
  lab/.style={font=\small}
]

\node[male]   (G1m) at (0,3.2) {};
\node[female] (G1f) at (1,3.2) {};
\draw[rel] (G1m)--(G1f);

\coordinate (sibmid1) at (0.5,2.75);
\draw[rel] (0.5,3.2) -- (sibmid1);
\draw[rel] (-3.6,2.75) -- (4.6,2.75);

\node[female] (D1) at (-3.6,2.2) {};
\node[female] (D2) at ( 0.5,2.2) {};
\node[female] (D3) at ( 4.6,2.2) {};

\draw[rel] (D1) -- (-3.6,2.75);
\draw[rel] (D2) -- ( 0.5,2.75);
\draw[rel] (D3) -- ( 4.6,2.75);

\node[male] (H1) at (-4.7,2.2) {};
\node[male] (H2) at (-0.6,2.2) {};
\node[male] (H3) at ( 3.5,2.2) {};
\draw[rel] (H1)--(D1);
\draw[rel] (H2)--(D2);
\draw[rel] (H3)--(D3);

\coordinate (cp12) at ($(H1)!0.5!(D1)$);
\coordinate (cp34) at ($(H2)!0.5!(D2)$);
\coordinate (cp56) at ($(H3)!0.5!(D3)$);


\coordinate (sibmid2) at ($(cp12)+(0,-0.55)$);
\draw[rel] (cp12) -- (sibmid2);
\draw[rel] ($(sibmid2)+(-0.8,0)$) -- ($(sibmid2)+(0.8,0)$);

\node[female] (p1) at ($(sibmid2)+(-0.8,-0.65)$) {};
\node[female] (p2) at ($(sibmid2)+( 0.8,-0.65)$) {};
\draw[rel] (p1) -- ($(sibmid2)+(-0.8,0)$);
\draw[rel] (p2) -- ($(sibmid2)+( 0.8,0)$);

\coordinate (sibmid3) at ($(cp34)+(0,-0.55)$);
\draw[rel] (cp34) -- (sibmid3);
\draw[rel] ($(sibmid3)+(-0.9,0)$) -- ($(sibmid3)+(0.9,0)$);

\node[female] (p3) at ($(sibmid3)+(-0.9,-0.65)$) {};
\node[female] (p4) at ($(sibmid3)+( 0.9,-0.65)$) {};
\draw[rel] (p3) -- ($(sibmid3)+(-0.9,0)$);
\draw[rel] (p4) -- ($(sibmid3)+( 0.9,0)$);

\coordinate (sibmid4) at ($(cp56)+(0,-0.55)$);
\draw[rel] (cp56) -- (sibmid4);
\draw[rel] ($(sibmid4)+(-0.8,0)$) -- ($(sibmid4)+(0.8,0)$);

\node[female] (p5) at ($(sibmid4)+(-0.8,-0.65)$) {};
\node[female] (p6) at ($(sibmid4)+( 0.8,-0.65)$) {};
\draw[rel] (p5) -- ($(sibmid4)+(-0.8,0)$);
\draw[rel] (p6) -- ($(sibmid4)+( 0.8,0)$);

\node[lab, below=2pt of p1] {$1$};
\node[lab, below=2pt of p2] {$2$};
\node[lab, below=2pt of p3] {$3$};
\node[lab, below=2pt of p4] {$4$};
\node[lab, below=2pt of p5] {$5$};
\node[lab, below=2pt of p6] {$6$};

\end{tikzpicture}
\caption{Pedigree with Generation I founders (one male, one female), Generation II with three daughters and their spouses, and Generation III with six female offspring (two per couple).}
\label{fig:pedigree_spouses_3daughters}
\end{figure}

\subsection*{Appendix G : Parameter estimates and precision using B-spline functions with 4 cubic polynomial basis functions}

\begin{table}[htbp]
\centering
\scriptsize
\setlength{\tabcolsep}{3pt}
\begin{tabular}{|c|cc|cc|cc|cc|cc|cc|cc|cc|}
\hline
 & \multicolumn{2}{c|}{$\gamma=0$}
 & \multicolumn{2}{c|}{$\gamma=0.001$}
 & \multicolumn{2}{c|}{$\gamma=0.005$}
 & \multicolumn{2}{c|}{$\gamma=0.01$}
 & \multicolumn{2}{c|}{$\gamma=0.05$}
 & \multicolumn{2}{c|}{$\gamma=0.1$}
 & \multicolumn{2}{c|}{$\gamma=0.5$}
 & \multicolumn{2}{c|}{$\gamma=1$} \\
\hline
Parameters
& Est. & s.e.
& Est. & s.e.
& Est. & s.e.
& Est. & s.e.
& Est. & s.e.
& Est. & s.e.
& Est. & s.e.
& Est. & s.e. \\
\hline
$\log(\alpha)$
& 1.079 & 0.042 & 1.079 & 0.042 & 1.079 & 0.042 & 1.079 & 0.042 & 1.079 & 0.042 & 1.079 & 0.041 & 1.078 & 0.041 & 1.078 & 0.041 \\

$\log(\lambda)$
& -13.690 & 0.743 & -13.692 & 0.742 & -13.695 & 0.741 & -13.695 & 0.740 & -13.701 & 0.738 & -13.707 & 0.735 & -13.721 & 0.729 & -13.720 & 0.727 \\

$\beta_{\text{gene}}$
& 1.939 & 0.214 & 1.940 & 0.213 & 1.940 & 0.213 & 1.941 & 0.213 & 1.942 & 0.213 & 1.942 & 0.213 & 1.942 & 0.214 & 1.941 & 0.214 \\

$b_1$
& -0.643 & 0.575 & -0.589 & 0.545 & -0.482 & 0.500 & -0.428 & 0.488 & -0.285 & 0.459 & -0.200 & 0.419 & 0.029 & 0.431 & 0.103 & 0.360 \\

$b_2$
& 3.050 & 1.891 & 2.787 & 1.603 & 2.267 & 1.027 & 2.021 & 0.876 & 1.520 & 0.822 & 1.278 & 0.634 & 0.655 & 0.417 & 0.449 & 0.349 \\

$b_3$
& -2.615 & 4.194 & -2.076 & 3.525 & -1.035 & 1.985 & -0.585 & 1.411 & 0.007 & 0.022 & 0.097 & 0.547 & 0.120 & 0.511 & 0.093 & 0.508 \\

$b_4$
& -0.246 & 5.171 & -0.787 & 4.605 & -1.830 & 3.330 & -2.248 & 2.805 & -2.399 & 1.782 & -2.079 & 1.367 & -0.979 & 0.803 & -0.613 & 0.782 \\

$h$
& 0.540 & 0.259 & 0.542 & 0.259 & 0.544 & 0.258 & 0.546 & 0.258 & 0.555 & 0.257 & 0.561 & 0.256 & 0.582 & 0.253 & 0.589 & 0.253 \\

\hline
$\sum_{j=1}^{4} b_j^2$
& 16.613 & --- & 13.041 & --- & 9.793 & --- & 9.664 & --- & 8.146 & --- & 6.004 & --- & 1.404 & --- & 0.597 & --- \\
\hline
\end{tabular}
\caption{Estimates and standard errors for the model parameters under different values of the smoothing parameter $\gamma$.}
\end{table}

\begin{figure}[H]
   \centering
 \includegraphics[scale=0.85]{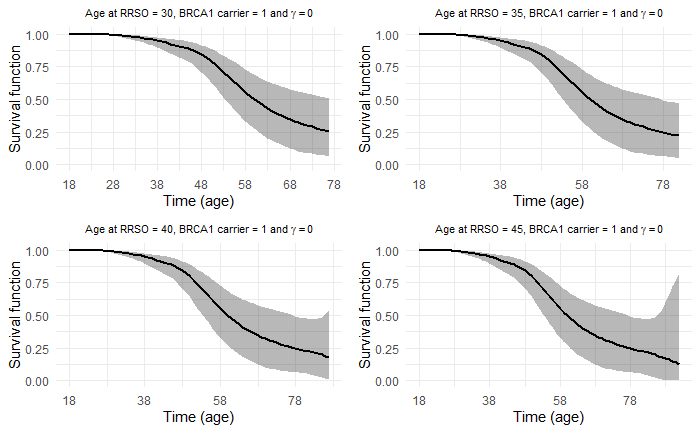}
 \caption{Estimate of $S(t|X_{ij})$, as a function of $t$, with $X_{ij1}$ equal to $30$, $35$,$40$ and $45$, and $X_{ij2}=1$ under  the smoothing parameter $\gamma=0$.  The confidence bands are obtained by the delta method and sandwich variance.}
\end{figure}

\begin{figure}[H]
   \centering
 \includegraphics[scale=0.85]{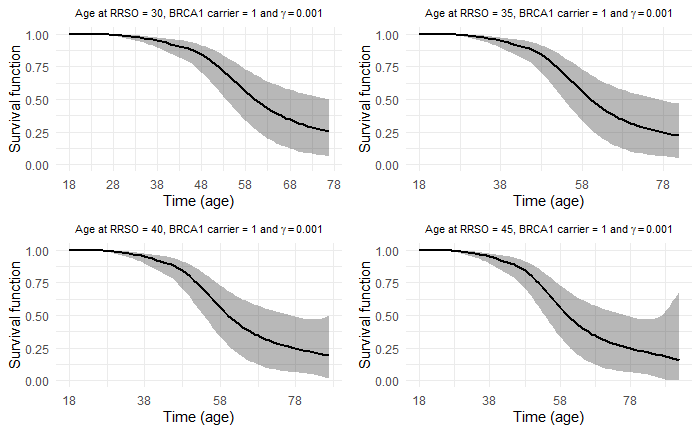}
 \caption{Estimate of $S(t|X_{ij})$, as a function of $t$, with $X_{ij1}$ equal to $30$, $35$,$40$ and $45$, and $X_{ij2}=1$ under  the smoothing parameter $\gamma=0.001$.  The confidence bands are obtained by the delta method and sandwich variance.}
\end{figure}

\begin{figure}[H]
   \centering
 \includegraphics[scale=0.85]{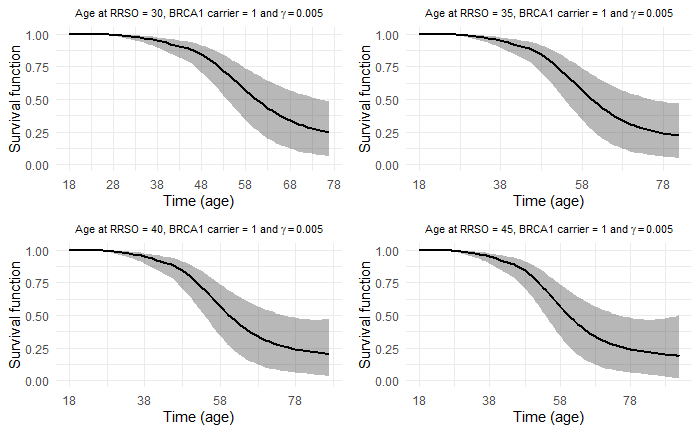}
 \caption{Estimate of $S(t|X_{ij})$, as a function of $t$, with $X_{ij1}$ equal to $30$, $35$,$40$ and $45$, and $X_{ij2}=1$ under  the smoothing parameter $\gamma=0.005$.  The confidence bands are obtained by the delta method and sandwich variance.}
\end{figure}

\begin{figure}[H]
   \centering
 \includegraphics[scale=0.85]{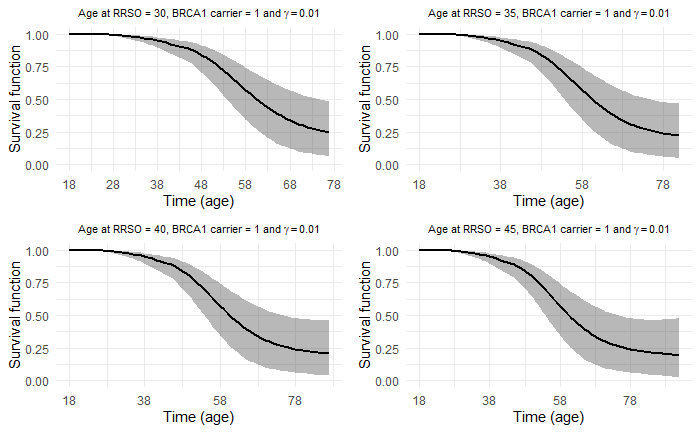}
 \caption{Estimate of $S(t|X_{ij})$, as a function of $t$, with $X_{ij1}$ equal to $30$, $35$,$40$ and $45$, and $X_{ij2}=1$ under  the smoothing parameter $\gamma=0.01$.  The confidence bands are obtained by the delta method and sandwich variance.}
\end{figure}

\begin{figure}[H]
   \centering
 \includegraphics[scale=0.85]{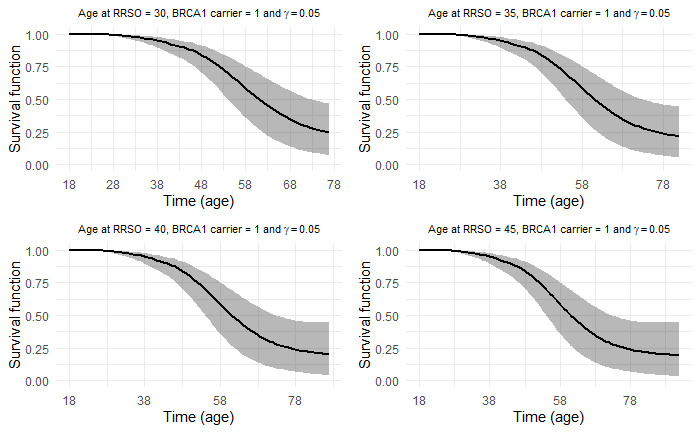}
 \caption{Estimate of $S(t|X_{ij})$, as a function of $t$, with $X_{ij1}$ equal to $30$, $35$,$40$ and $45$, and $X_{ij2}=1$ under  the smoothing parameter $\gamma=0.05$.  The confidence bands are obtained by the delta method and sandwich variance.}
\end{figure}

\begin{figure}[H]
   \centering
 \includegraphics[scale=0.85]{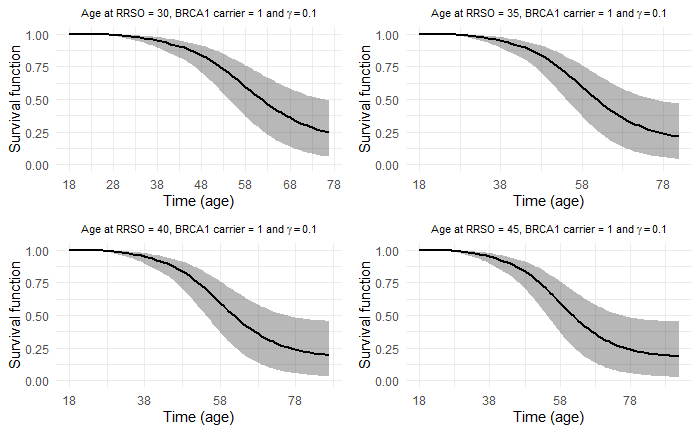}
 \caption{Estimate of $S(t|X_{ij})$, as a function of $t$, with $X_{ij1}$ equal to $30$, $35$,$40$ and $45$, and $X_{ij2}=1$ under  the smoothing parameter $\gamma=0.1$.  The confidence bands are obtained by the delta method and sandwich variance.}
\end{figure}

\begin{figure}[H]
   \centering
 \includegraphics[scale=0.85]{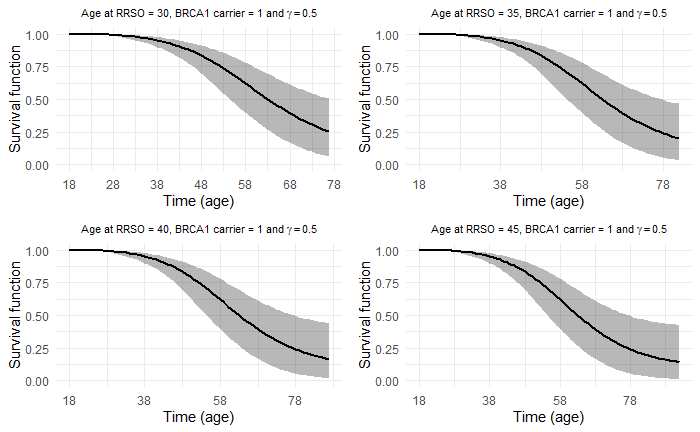}
 \caption{Estimate of $S(t|X_{ij})$, as a function of $t$, with $X_{ij1}$ equal to $30$, $35$,$40$ and $45$, and $X_{ij2}=1$ under  the smoothing parameter $\gamma=0.5$.  The confidence bands are obtained by the delta method and sandwich variance.}
\end{figure}

\begin{figure}[H]
   \centering
 \includegraphics[scale=0.85]{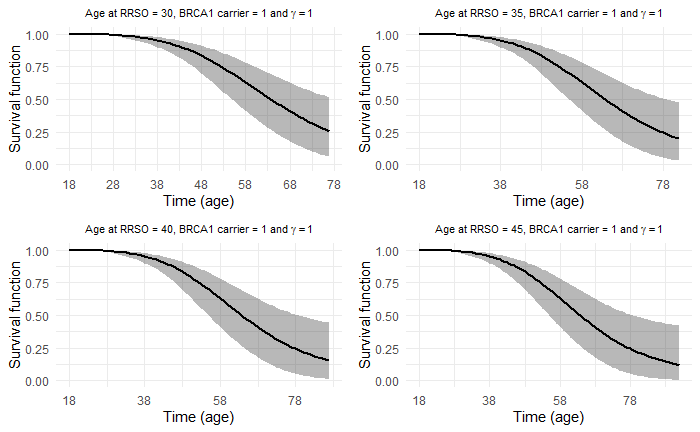}
 \caption{Estimate of $S(t|X_{ij})$, as a function of $t$, with $X_{ij1}$ equal to $30$, $35$,$40$ and $45$, and $X_{ij2}=1$ under  the smoothing parameter $\gamma=1$.  The confidence bands are obtained by the delta method and sandwich variance.}
\end{figure}